\documentclass[%
 pdftex,
 11pt,
 aps,
 prd,
 floatfix,
 showpacs,
 nofootinbib,
 tightenlines,
 amsmath
]{revtex4-1}

\usepackage{diagbox}
\usepackage[figuresright]{rotating}
\usepackage{graphicx}
\usepackage{cmap} 
\usepackage{bm}
\usepackage[dvipsnames]{xcolor}
\usepackage{hyperref}
\hypersetup{%
colorlinks=true,%
citecolor=Blue,%
filecolor=Black,%
linkcolor=Blue,%
urlcolor=Violet
}
\usepackage[activate={true,nocompatibility},final,tracking=true,kerning=true,spacing=true,factor=1100,stretch=10,shrink=10]{microtype}

\graphicspath{{./figures/}}

\newcommand{\mara}{{MARATHON}}
\newcommand{\bonus}{{BoNuS}}

\newcommand{\ud}  {\mathrm{d}}
\newcommand{\gev} {\ \mathrm{GeV}}
\newcommand{\mev} {\ \mathrm{MeV}}
\newcommand{\gevsq}{\ \mathrm{GeV}^2}

\newcommand{\ceps}{\varepsilon}

\newcommand{\eq}[1]{Eq.~(\ref{#1})}
\newcommand{\Eqs}[2]{Eq.~(\ref{#1}) and (\ref{#2})}
\newcommand{\eqs}[1]{Eqs.~(\ref{#1})}
\newcommand{\ptsq}{p_{\perp}^2}
\newcommand{\updtxt}[1]{#1}

\AtBeginDocument{%
    \newwrite\bibnotes
    \def\bibnotesext{Notes.bib}
    \immediate\openout\bibnotes=\jobname\bibnotesext
    \immediate\write\bibnotes{@CONTROL{REVTEX41Control}}
    \immediate\write\bibnotes{@CONTROL{%
    apsrev41Control,author="08",editor="0",pages="0",title="0",year="1"}}
     \if@filesw
     \immediate\write\@auxout{\string\citation{apsrev41Control}}%
    \fi
}%
\begin{document}

\title{%
Nuclear effects in the deuteron and global QCD analyses
}

\author{S.~I.~Alekhin}
\email[]{sergey.alekhin@desy.de}
\affiliation{II. Institut f\"ur Theoretische Physik, Universit\"at Hamburg, \\
  Luruper Chaussee 149, D--22761 Hamburg, Germany}
\author{S.~A.~Kulagin}
\email[]{kulagin.physics@gmail.com}
\affiliation{Institute for Nuclear Research of the Russian Academy of Sciences, \\
117312 Moscow, Russia}
\author{R.~Petti}
\email[]{roberto.petti@cern.ch}
\affiliation{Department of Physics and Astronomy, \\
University of South Carolina, Columbia, South Carolina 29208, USA}


\begin{abstract}
\noindent
We report the results of a new global QCD analysis, which includes deep-inelastic $e/\mu$
scattering  data off proton and deuterium, as well as Drell-Yan lepton pair production
in proton-proton and proton-deuterium collisions and $W^\pm/Z$ boson production data from
$pp$ and $p \bar p$ collisions at the LHC and Tevatron.
Nuclear effects in the deuteron are treated in terms of
\updtxt{%
a nuclear convolution approach with bound off-shell nucleons within a weak binding approximation.
The off-shell correction is controlled by a universal function of the Bjorken variable $x$ describing the modification}
of parton distributions in bound nucleons, which is determined in our analysis
along with the parton distribution functions of the proton.
A number of systematic studies are performed to estimate the uncertainties arising
from the use of various deuterium datasets, from the modeling of higher twist contributions to the
structure functions, from the treatment of target mass corrections, as well as from the nuclear corrections in the deuteron.
We obtain predictions for the ratios $F_2^n/F_2^p$, and $d/u$, focusing on the region of large $x$.
We also compare our results with the ones obtained by other QCD analyses,
as well as with the recent data from the \mara{} experiment.

\end{abstract}


\maketitle

\section{Introduction}
\label{sec:intro}

An accurate determination of the parton distribution functions (PDFs) in the proton and the neutron
is of primary importance for modern high-energy physics, as PDFs determine the leading contribution to
the cross sections of various high-energy processes.
Since PDFs are not directly observable, they are usually extracted phenomenologically from global QCD analyses
to experimental data at large momentum transfer, including  lepton deep inelastic scattering (DIS),
lepton-pair production (Drell-Yan process), jet production, and $W^\pm/Z$ boson production in hadron collisions
(for a review see, e.g., Ref.~\cite{Accardi:2016ndt}).
While the abundant data available from a hydrogen target allow a reliable determination of the PDF content of the proton,
data from various nuclei---most notably deuterium---are required as effective neutron targets to constrain the
parton content of the neutron. Furthermore, a combination of hydrogen and deuterium data has been commonly
used to separate the $u$ and $d$ quark PDFs, in particular, at large values of Bjorken $x$.%
\footnote{%
For a recent discussion of the impact of deuterium data on global QCD analyses see Refs.~\cite{Accardi:2021ysh,Ball:2020xqw}.}

Precision studies require one to address the effects of nuclear environment at the parton level.
While the nuclear effects in nuclear PDFs (nPDFs) analyses are usually treated empirically~\cite{Hirai:2001np,deFlorian:2011fp,Kovarik:2015cma,Eskola:2021nhw},
a number of physics mechanisms are known to affect the PDFs and the structure functions (SFs) of the bound nucleons (for a review see, e.g.,~\cite{Arneodo:1992wf,Geesaman:1995yd,Norton:2003cb}).
In the region of large $x$, the relevant nuclear effects are related to the smearing of the cross sections
with the nuclear momentum distribution~\cite{Atwood:1972zp}
(Fermi motion), together with the nuclear binding correction~\cite{Akulinichev:1985ij}.
In addition
\updtxt{to these corrections, which have kinematical origin},
nuclear effects related to the dynamical modification of the internal parton structure have to be addressed in bound nucleons.
In Refs.~\cite{Kulagin:1994fz,Kulagin:2004ie}, such a modification is related to the off-mass-shell effect, i.e., the dependence of bound nucleon SFs on its virtual mass squared $p^2=p_0^2-\bm p^2$, where $p_0$ and $\bm p$ are the nucleon energy and momentum, respectively.
This dependence is treated in Refs.~\cite{Kulagin:1994fz,Kulagin:2004ie} as a perturbative correction
in the nucleon virtuality $v=(p^2-M^2)/M^2$, relying on the fact that the typical nucleon momentum and energy are small compared to the nucleon mass $M$ in the nuclear ground state.
Within this weak binding approximation, the corresponding nuclear correction is controlled
by the SFs' derivative with respect to $p^2$, which is described in Ref.~\cite{Kulagin:2004ie} in terms of
a dimensionless function $\delta f(x)$.
Additional effects related to the meson-exchange currents and the nuclear shadowing are relevant at intermediate and small $x$ values.
A model combining all of these effects has been successfully used to quantitatively explain the
observed dependencies on $x$, invariant momentum transfer squared $Q^2$, nuclear mass number $A$ of the
nuclear DIS data in a wide range of targets from ${}^3$He to ${}^{207}$Pb~\cite{Kulagin:2004ie,Kulagin:2007ju,Kulagin:2010gd}.
The same model also demonstrates an excellent agreement with
the magnitude, the $x$ and mass dependence of the nuclear Drell-Yan (DY) data~\cite{Kulagin:2014vsa}, as well as with the data on
the differential cross sections for $W^\pm/Z$ boson production in proton-lead
collisions at the LHC~\cite{Ru:2016wfx}.

The off-shell effect is an important contribution to the
full nuclear correction.
The corresponding function $\delta f$
was determined for the isoscalar nucleon
from an analysis of nuclear DIS data on the cross-section ratios $\sigma^A/\sigma^d$~\cite{Kulagin:2004ie}.
The function $\delta f(x)$ was also \updtxt{independently} extracted
together with the proton PDFs in
global QCD analyses of proton and deuterium DIS data~\cite{Alekhin:2017fpf,Accardi:2016qay}.
The results of Ref.~\cite{Alekhin:2017fpf} on $\delta f$ are consistent with the previous determination from
nuclei with $A\geq 4$~\cite{Kulagin:2004ie}.
However, Refs.~\cite{Accardi:2016qay} and~\cite{Alekhin:2017fpf}
strongly disagree on both the values of the function $\delta f$ and on the ratio of the $d$ and $u$ quark PDFs at large $x$.
These observations motivate the present study, in which we perform a new global QCD analysis with
updated sets of deuterium DIS data. We discuss a number of systematic studies aimed at understanding
the uncertainties associated with a number of effects, including the consistency of various deuterium datasets,
the treatment of target mass correction, and the modeling of higher-twist contributions and of the nuclear corrections in the deuterium.
We also provide our predictions on the ratios
$F_2^n/F_2^p$ and $d/u$
and compare them with the ones from the QCD analyses of Refs.~\cite{Accardi:2016qay,Hou:2019efy,Cridge:2021qfd,NNPDF:2021njg},
as well as with the recent data from \mara{} experiment~\cite{MARATHON:2021vqu}.

The paper is organized as follows.
In Sec.~\ref{sec:frmwrk}, we outline the theory framework used in our analysis of the proton and deuterium DIS.
In Sec.~\ref{sec:analysis}, we discuss the data samples and the details of our analysis.
In Sec.~\ref{sec:results}, we summarize our results,
while in Sec.~\ref{sec:unc}, we discuss the
uncertainties associated with the use of different
deuterium datasets and with the
modeling of the structure functions.
In Sec.~\ref{sec:discus}, we compare our predictions
on $F_2^n/F_2^p$ with \mara{} data and with
the results of other QCD analyses, including the $d/u$ ratio.
In Appendix~\ref{sec:pulls}, we show the pulls obtained in our fit from different deuterium datasets.
In Appendix~\ref{sec:psi}, we discuss in detail the phase space in the nuclear convolution equations employed in our analysis.

\section{Theory framework}
\label{sec:frmwrk}

\subsection{Nucleon structure functions}
\label{sec:pn}

The inclusive spin-independent electron(muon)-nucleon inelastic cross section
is described by two SFs, $F_T=2xF_1$ and $F_2$,
which depend on two independent variables, the invariant momentum transfer squared $Q^2=-q^2$
and the dimensionless Bjorken $x=Q^2/(2p\cdot q)$, where $p$ is the nucleon four-momentum and $q$ is the four-momentum transfer.

A common framework to describe
the DIS is the operator product expansion (OPE),
which introduces the power series in $Q^{-2}$ (twist expansion).
To the first order, i.e., in the leading twist (LT), the SFs are fully determined by the PDFs.
Corrections from the higher-twist (HT) quark-gluon operators should also be supplemented
by those arising from the finite nucleon mass (target mass correction, or TMC)~\cite{Georgi:1976ve}.
We also note that for the sake of computing the nuclear SFs (see Sec.~\ref{sec:d}),
the nucleon SFs are required in the off-mass-shell region $p^2<M^2$,
where $M$ is the nucleon mass.
The unpolarized nucleon SFs in the DIS region can then be written as follows
\begin{equation}\label{eq:sfdis}
F_i(x,Q^2,p^2) = F_i^{\text{TMC}}(x,Q^2,p^2) + H_i/ Q^2,
\end{equation}
where $i=T,2$ and $F_i^{\text{TMC}}$ are the corresponding LT SFs
corrected for the target mass effect and $H_i$
describe the dynamical twist-4 contribution
(for brevity, we suppress explicit notation to the twists higher than four).
In this study, we consider two different phenomenological HT models:
(1) the additive HT model, in which we assume $H_i=H_i(x)$ and
(2) the multiplicative HT model~\cite{Virchaux:1991jc},
in which $H_i$ is assumed to be proportional to the corresponding LT SF, $H_i=F_i^\text{LT}(x,Q^2) h_i(x)$.
The HT terms from both models are addressed in this study.

The LT SFs are computed using the nucleon PDFs and coefficient functions,
which are subject to a power series in the QCD coupling constant.
The neutron LT SFs are computed in terms of the proton PDFs relying on the isospin symmetry
of $u$ and $d$ quark PDFs.
The isospin relations for the HT terms are not so obvious.
By default, we assume $H_i^p=H_i^n$.%
\footnote{%
We note, however, that a nonzero isovector component $H_2^p-H_2^n$ was obtained
in a QCD fit~\cite{Alekhin:2003qq}, although with rather large fit uncertainties.
The difference $H_T^p-H_T^n$ was consistent with 0 within uncertainties.}
We also consider the relation  $h_i^p=h_i^n$  with the multiplicative HT model.

To account for the TMC, we follow the Georgi-Politzer OPE approach~\cite{Georgi:1976ve}.
Since the calculation of the nuclear SFs requires the nucleon SFs in the off-shell mass region,
we analytically continue the equations of Ref.~\cite{Georgi:1976ve} into the off-shell region
by replacing the nucleon mass squared $M^2$ with $p^2$. We have
\begin{subequations}\label{eq:TMC}
\begin{align}
\label{eq:TMC:T}
	F_T^{\text{TMC}}(x,Q^2,p^2) = &
\frac{x^2}{\xi^2 \gamma} F_T^\text{LT}(\xi,Q^2,p^2) +
\frac{2x^3 p^2}{Q^2\gamma^2}\int^1_\xi \frac{d u}{u^2}
 F_2^\text{LT}(u,Q^2,p^2), 
\\
\label{eq:TMC:2}
	F_2^{\text{TMC}}(x,Q^2,p^2) = &
	\frac{x^2}{\xi^2 \gamma^3} F_2^\text{LT}(\xi,Q^2,p^2) +
	\frac{6x^3p^2}{Q^2\gamma^4}\int^1_\xi \frac{d u}{u^2}
 F_2^\text{LT}(u,Q^2,p^2), 
\end{align}
\end{subequations}
where $\xi=2x/(1+\gamma)$ is the Nachtmann variable, and $\gamma=(1+4x^2 p^2/Q^2)^{1/2}$.
Note that in \eqs{eq:TMC} we drop the terms of order $x^4p^4/Q^4$, which produce numerically small contributions in the considered region.
It should be commented that \Eqs{eq:TMC:T}{eq:TMC:2} lead to a nonzero SFs at $x\to1$.
However, in practice this violation of the inelastic threshold behavior does not
affect the DIS region, which is characterized by high values of the invariant mass $W$ of the produced hadronic states.

In the off-mass-shell region, the SFs explicitly depend on the nucleon invariant mass squared $p^2$.
This dependence has two different sources:
(i) the terms $p^2/Q^2$ in \eqs{eq:TMC}, which lead to power terms at large values of $Q^2$
and
(ii) nonpower terms from the off-shell dependence of the LT SFs.
Following Refs.~\cite{Kulagin:1994fz,Kulagin:2004ie}, we note that for computing the nuclear SFs,
it would be sufficient to know the proton and the neutron SFs in the vicinity of the mass shell $p^2=M^2$.
We then treat the nucleon virtuality $v=(p^2-M^2)/M^2$ as a small parameter and expand SFs in series in $v$.
To the leading order in $v$, we have
\begin{align}
	\label{SF:OS}
	F_i^\text{LT}(x,Q^2,p^2) &=
	F_i^\text{LT}(x,Q^2,M^2)\left[ 1+\delta f_i(x,Q^2)\,v \right],
	\\
	\label{delf}
	\delta f_i(x,Q^2) &= M^2\partial_{p^2}\ln F_i^\text{LT}(x,Q^2,p^2),  
\end{align}
where $F_i^\text{LT}$ on the right-hand side in \eq{SF:OS} are the structure
functions $i=T,2$ of the on-mass-shell nucleon, and $\partial_{p^2}$ in \eq{delf} denotes
the partial derivative with respect to $p^2$ taken on the mass shell $p^2=M^2$.
According to \eq{delf}, the function $\delta f_i$ describes the relative
modification of the nucleon LT $F_i$ in the vicinity of the mass shell,
which is related to the corresponding PDF modification.

In this study, we assume the function $\delta f$ to be the same for $F_T$ and $F_2$
motivated by the fact that $F_T\approx F_2$ in the region of large $x$.
The function $\delta f$ drives the nuclear correction associated with the modification of
the bound nucleon in the nuclear environment~\cite{Kulagin:2004ie}.
Detailed studies of nuclear DIS, DY production of the lepton pair and $W/Z$ boson in Refs.\cite{Kulagin:2004ie,Kulagin:2010gd,Kulagin:2014vsa,Ru:2016wfx,Alekhin:2017fpf}
are consistent with no significant scale  and nucleon isospin dependencies of $\delta f$.
We thus assume the same $\delta f=\delta f(x)$ function for the proton and the neutron.

Note that \eq{SF:OS} holds in the vicinity of the mass shell where $|v|\ll 1$.
In computing the nuclear SFs, we integrate over the bound nucleon momentum as discussed in Sec.~\ref{sec:d}.
For kinematics reason, $p^2<M^2$ and $v<0$ for bound nucleons.
Using the results of Ref.~\cite{Kulagin:2004ie}, we have $\delta f\sim 1$ at large $x>0.6$.
Then the off-shell correction in \eq{SF:OS} is large and negative for $v\sim -1$,
and the off-shell SFs may be negative in this region.
Since the values $|v|\gtrsim 1$
are outside of the region of applicability of the linear approximation in $v$, \eq{SF:OS},
we consider the following model in the full region of $v$:
\begin{equation}
\label{SF:OS2}
F_i^\text{LT}(x,Q^2,p^2) = F_i^\text{LT}(x,Q^2,M^2)\exp[\delta f(x) v] .
\end{equation}
This equation ensures the positivity of SFs in the off-shell region, and
for a small off-shell correction \eq{SF:OS2} is identical to \eq{SF:OS}.
In the study of the deuteron SFs, we consider both \eq{SF:OS} and \eq{SF:OS2}.

\subsection{Deuteron structure functions}
\label{sec:d}

We assume that the nuclear DIS in the region $x>0.1$
is dominated by the incoherent scattering off the bound protons and neutrons and
consider the process in the target rest frame.
The deuteron structure functions  can be written as follows~\cite{Alekhin:2003qq,Kulagin:2004ie}:
\begin{equation}
	\label{eq:IA}
	F_i^{d}(x,Q^2) = \int \ud^3\bm p \left|\Psi_d(\bm p)\right|^2 K_{ij}
	\left[ F_j^p(x',Q^2,p^2)+F_j^n(x',Q^2,p^2) \right],
\end{equation}
where $i,j=T,2$, and we assume a summation over the repeated subscript $j$.
The integration is performed over the bound nucleon momentum $\bm p$,
and $\Psi_d(\bm p)$ is the deuteron wave function in the momentum space,
which is normalized as
\begin{equation}\label{eq:wfnorm}
 \int\ud^3\bm p \left|\Psi_d(\bm p)\right|^2 = 1.
\end{equation}
Because of the energy-momentum conservation, the four-momentum of the struck proton (neutron) is
$p=(M_d-\sqrt{M^2+\bm p^2}, \bm{p})$, where $M_d$ is the deuteron mass, and $M$ is the mass of
residual nucleon [$M=M_n$ for the proton contribution and $M=M_p$ for the neutron contribution in \eq{eq:IA}].
We use a coordinate system in which the momentum transfer $\bm{q}$ is antiparallel to the $z$ axis, and
$p_z$ and $\bm{p}_\perp$ are the longitudinal and transverse component of the nucleon momentum,
$p^2=p_0^2-\bm p^2$ and $x'=Q^2/(2p\cdot q)$ are the invariant mass
and the Bjorken variable of the off-shell nucleon, respectively.
The kinematic factors $K_{ij}$ are~\cite{Kulagin:2004ie}
\begin{subequations}\label{eq:Kij}
	\begin{align}
	K_{TT} &= \left(1+\frac{\gamma p_z}{M}\right), &K_{T2} &= 2\frac{{x'}^2\bm p_\perp^2}{Q^2},
\\
	K_{2T} &= 0, &K_{22} &= \left(1+\frac{\gamma p_z}{M}\right)\left(1+\frac{{x'}^2(4p^2+6\bm{p}_\perp^2)}{Q^2}\right) \frac{1}{\gamma^2},
	\end{align}
\end{subequations}
where $\gamma=(1+4x^2 M^2/Q^2)^{1/2}$.
Note that \Eqs{eq:IA}{eq:Kij} are the result of a series expansion of
relativistically covariant operators in the parameters $\bm p/M$ and $(p_0-M)/M$
to order $\bm p^2/M^2$
(for more detail, see \cite{Kulagin:1994fz,Kulagin:2004ie} and Appendices B and C of Ref.~\cite{Kulagin:2008fm}).
The factor $1+\gamma p_z/M$ in \eq{eq:Kij} describes the change in the virtual photon flux for a bound nucleon with the momentum $\bm p$ compared to the corresponding flux for the nucleus at rest.
Note also the term $K_{T2}\sim x^2\bm p_\perp^2/Q^2$ resulting from a mixing effect between the longitudinal and transverse structure functions at finite values of $Q^2$, which is due to the transverse motion of the bound nucleon.

Assuming no $p^2$ dependence of the nucleon structure functions,
in the limit $Q\gg M$, \eq{eq:IA} reduces to the standard convolution of the nucleon SFs
with the nucleon distribution over the light-cone momentum $y=(p_0+p_z)/M$ in the deuteron.
In the presence of an off-shell $p^2$ dependence we have a generalized convolution,
which involves the integration over the light-cone momentum $y$ and the nucleon virtuality $p^2$~\cite{Kulagin:1994fz}.
The phase space at finite $Q^2$ used in \eq{eq:IA} is discussed in more detail in Appendix~\ref{sec:psi}.

In the region $x<0.1$, the corrections due to the meson-exchange currents and the nuclear shadowing, at even smaller values of $x\ll0.1$, are relevant.
In this study, while focusing on $x>0.1$, we treat these effects following Refs.~\cite{Kulagin:2004ie,Kulagin:2014vsa}.

\section{Off-shell Function within Global QCD Analysis}\label{sec:analysis}

\subsection{Data samples}
\label{sec:data}

\begin{table}[htb!]
\begin{center}
\caption{\label{tab:data}
The list of DIS data on the deuterium target employed in the present analysis alongside with the values of $\chi^2/\text{NDP}$ and normalization factors obtained in the fit in comparison with the experimentally determined normalization errors.}
\begin{tabular}{l|c|c|c|c|c|c|c|c} \hline
Facility & Experiment & Reference & Beam  & Beam energy& Observable & Normalization & Normalization & \raisebox{0pt}{$\frac{\chi^2}{\text{NDP}}$} \\
&           &           &       & (GeV)      &            & factor        & error(s) (\%) & \\
\hline\hline
SLAC& E49a  & \cite{Bodek:1979rx,Whitlow:1990gk} & $e$  & $11\div 19.5$  & $\frac{\ud^2\sigma^d}{\ud E^\prime \ud\Omega}$ &0.988(10) & 2.1\footnote{A general normalization uncertainty for the SLAC experiments derived from re-analysis of those data. The contributions of marginal size also apply to particular datasets~\cite{Whitlow:1990dr}.}&25/59 \\
"& E49b & " & "  &$4.5\div 18$ & "  &0.996(10) &"& 187/145 \\
"& E87  & " & "  &$8.7\div 20$ & "  &1.000(9) &"& 114/109 \\
"& E89b & \cite{Mestayer:1982ba,Whitlow:1990gk} & "  &$10.4\div 19.5$ & " &0.987(9)  &"& 52/72 \\
"& E139 & \cite{Gomez:1993ri,Whitlow:1990gk} & "  &$8\div 24.5$& " &1.002(9)&"& 8/17 \\
"& E140 & \cite{Dasu:1993vk,Whitlow:1990gk} & "  &$3.7\div 19.5$& "  & 1&1.7&  25/26 \\
CERN& BCDMS  & \cite{Benvenuti:1989fm} & $\mu$  &$100\div 280$ & $\frac{\ud^2\sigma^d}{\ud x\ud Q^2}$ & 0.989(7) &3 & 273/254 \\
" &NMC  & \cite{Arneodo:1996kd}  & "  & $90\div 280$ & $F_2^d/F_2^p$ & 1 &$<0.15$& 155/165 \\
DESY &HERMES  & \cite{Airapetian:2011nu} & $e$  & 27.6 & $\sigma^d/\sigma^p$ & 1 &1.4& 21/30 \\
JLab &E00-116 & \cite{Malace:2009kw} &  " &5.5&  $\frac{\ud^2\sigma^d}{\ud E^\prime \ud\Omega}$ & 0.981(10) &1.75& 208/136 \\
" &BoNuS & \cite{Tkachenko:2014byy} &  " &4.2, 5.2& $F_2^n/F_2^d$ &0.97(9)  &$7\div 10$& 90/63 \\
" &MARATHON & \cite{MARATHON:2021vqu}&  " &10.6& $\sigma^d/\sigma^p$ & 1 &0.55& 8/7\\
\hline
     Total     & &  &  &  &  &  & &1166/1083 \\
\hline
\end{tabular}
\end{center}
\end{table}

\begin{figure}[htb]
\centering
\includegraphics[width=1.0\textwidth]{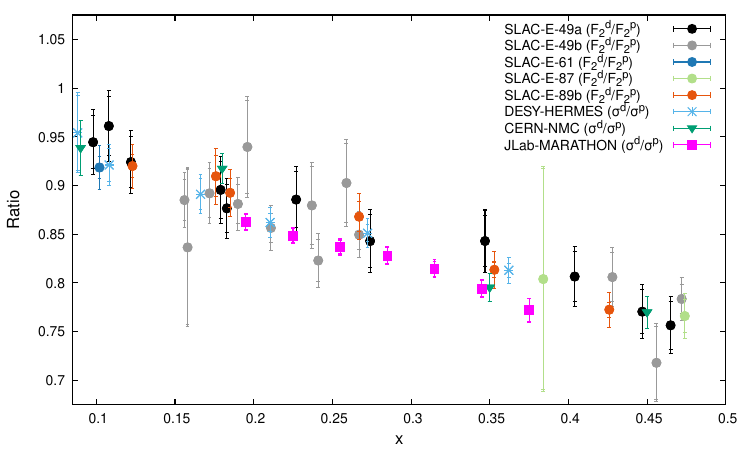}
\caption{%
Data on $F_2^d/F_2^p$ (or $\sigma^d/\sigma^p)$ for $0.1<x<0.5$ from various experiments. Data legend is shown in the plot. The data points were selected within interval $Q^2=14x\,(\mathrm{GeV}^2)\pm10\%$
to facilitate the comparison with Ref.~\cite{MARATHON:2021vqu}. The inner error bar shows the statistical and systematic error taken in quadrature, while the outer error bar in addition includes the normalization error listed in Table~\ref{tab:data}.
}
\label{fig:dp14x}
\end{figure}

The present study is an update of our former analysis~\cite{Alekhin:2017fpf}
based on the data on the DIS of charged leptons off hydrogen and deuterium
combined with the ones on $W^\pm/Z$ boson production at hadron
colliders. The latter samples allow the separation of the $u$ and $d$ quark
distributions in a wide range of $x$ that, in turn, provides a basis
for studying nuclear effects in the deuteron for the DIS structure functions.
The deuterium datasets employed for this purpose are listed in
Table~\ref{tab:data}. They comprise the ones used in the
analysis of Ref.~\cite{Alekhin:2017fpf} supplemented by
the most recent results on $\sigma^d/\sigma^p$
by the \mara{} experiment at Jlab~\cite{MARATHON:2021vqu}.
Due to the increased energy of the upgraded Jlab beam,
the \mara{} data cover a much wider kinematics as compared to the earlier JLab \bonus{} experiment~\cite{Tkachenko:2014byy}.%
\footnote{Note that while the \mara{} nuclear data covers the region $0.19<x<0.85$, the measurement of the ratio $\sigma^d/\sigma^p$ is available for a limited region $0.19<x<0.4$.}
Besides, a dedicated study performed
by \mara{} allowed one to reduce the normalization uncertainty in its
measurements to unprecedented level of 0.55\%. This guarantees a superior
statistical significance of the \mara{} data over both the
original \bonus{} sample~\cite{Tkachenko:2014byy} and the results of the
study~\cite{Griffioen:2015hxa} based on the \bonus{} measurements.
The results on $F_2^d/F_2^N$ derived in Ref.~\cite{Griffioen:2015hxa}
from the \bonus{} data on $F_2^n/F_2^d$ using a parametrization of
$F_2^p$ were employed in our earlier study~\cite{Alekhin:2017fpf}.
However, since they are sensitive to model assumptions about the $F_2^p$ shape,
in the present study, we select the original \bonus{}
data in order to reduce the
model dependence of the analysis.
To provide a complete representation of the relevant data,
we also add to the fit the DIS data collected in the
Jlab-E00-116~\cite{Malace:2009kw} and DESY-HERMES~\cite{Airapetian:2011nu}
experiments. Finally, we replace the deuteron NMC data~\cite{Arneodo:1996qe}
used in Ref.~\cite{Alekhin:2017fpf} by more recent measurements~\cite{Arneodo:1996kd}
performed with a higher luminosity.
In Fig.~\ref{fig:dp14x}, we illustrate the accuracy of various datasets by plotting the ratio $F_2^d/F_2^p$ in the range $0.1<x<0.5$ and for $Q^2=14x\,(\mathrm{GeV}^2)\pm10\%$ from various measurements~\cite{Whitlow:1990dr,Arneodo:1996kd,Airapetian:2011nu,MARATHON:2021vqu}.%
\footnote{We note that the reanalysed SLAC data~\cite{Whitlow:1990dr}
given in Fig.~\ref{fig:dp14x} are somewhat different from the original
data~\cite{Bodek:1979rx} due to updated radiative corrections and $x$-rebinning~\cite{Petratos:pc}.
The normalization of the original SLAC data is more consistent with the \mara{} $F_2^d/F_2^p$ data, as shown in Ref.~\cite{MARATHON:2021vqu}.}
This selection of the $Q^2$ band is motivated by kinematics of the \mara{} experiment~\cite{MARATHON:2021vqu}.

Whenever possible, we select the data on the cross sections and their
ratios rather than on the structure functions $F_2$. This makes the modeling more involved,
however, allows for a consistent account of the contribution from the
structure function $F_L$, since the various experiments do not follow
a common convention on the shape of $F_L$ when extracting $F_2$ from
the cross-section measurements. As an exception, the
\bonus{}~\cite{Tkachenko:2014byy} and the NMC~\cite{Arneodo:1996kd} data
are taken in the form of the ratios $F_2^n/F_2^d$  and $F_2^d/F_2^p$,
respectively, as the cross-section results have not been released
by these experiments. Such an inconsistency can be, however,
justified since the $F_L$ contribution to great extent cancels out in the ratios.
To ensure a perturbative QCD description of the leading-twist terms in the
DIS structure functions,
we impose a general cut of $Q^2>2.5\gevsq$ and $W^2>3\gevsq$.
For the \bonus{} data~\cite{Griffioen:2015hxa} used in our earlier
study~\cite{Alekhin:2017fpf}, a relaxed cut of $Q^2>1.5\gevsq$ was selected
in order to increase the statistical significance of this sample.
In the present analysis, which includes the precision \mara{} data, this exemption
is not applied and the \bonus{} data~\cite{Tkachenko:2014byy} are considered
within a common framework.

Information about the point-to-point correlation of systematic errors in
the data is taken
into account in the fit whenever available. In particular, a detailed breakdown of the
systematic uncertainties over independent sources is provided for
the SLAC, CERN-BCDMS, CERN-NMC, and JLab-\bonus{} experiments. For the Jlab-E00-116 and
\mara{} datasets, only the overall systematic uncertainty is published, and
in the present fit, it is combined in quadrature
with the statistical (uncorrelated) uncertainties.
The systematic uncertainties of DESY-HERMES measurements are separated by
sources; however, no information about their point-to-point correlation
was provided.
For this reason, we select for our fit the DESY-HERMES data on the ratio $\sigma^d/\sigma^p$,
where the correlated uncertainties partially cancel. The remaining systematic uncertainty,
except of the normalization one, are
combined with the statistical uncertainty, in line with the approximation adopted
in the DESY-HERMES analysis of their own data~\cite{Airapetian:2011nu}.

The normalization uncertainty,
a peculiar case of systematic errors,
often dominates the uncertainty of the datasets considered. Furthermore,
the normalization factors for the available SLAC-E49a, E49b, E87, E89b, E139, and CERN-NMC
datasets were estimated by comparing them to
the measurements of
the SLAC-E140 experiment with the normalization uncertainty of $1.7\%$.
Following a similar approach, we release the normalization
factors of those data and determine such factors from a fit
simultaneously with other parameters.
Furthermore, this procedure is also applied to the CERN-BCDMS and
Jlab-E00-116 data allowing for improvement of their instrumentally determined normalizations.
For the CERN-NMC and DESY-HERMES data, which were taken in the form of ratios,
the impact of the normalization
uncertainty is greatly reduced; therefore, their normalizations were kept fixed.
A similar treatment was applied to the \mara{} data, which have
a very accurate luminosity monitoring and for which we also avoid normalization
tuning. This allows us to use those data
for the calibration of the other datasets, in addition to the SLAC-E140 data set.

\subsection{Analysis setup}
\label{sec:fit}

The leading-twist PDFs,  which are necessary for
computing the nucleon SFs Eq.~(\ref{eq:sfdis}), are parametrized
using the shape employed in the ABMP16 fit~\cite{Alekhin:2017kpj} and
in our earlier analysis~\cite{Alekhin:2017fpf}.
The DIS SFs are treated as outlined in Sec.~\ref{sec:pn}.
The functions $H_T(x)$ and $H_2(x)$, which describe the HT contributions, are treated independently and are
parametrized in a model-independent form of spline polynomials interpolating between the points
$x=(0, 0.1, 0.3, 0.5, 0.7, 0.9, 1)$
with the values (HT coefficients) determined on this grid.
We assume the HT terms to be independent of the nucleon isospin state.
The $Q^2$ dependence of the LT component of the nucleon SFs was computed taking into account NNLO perturbative QCD (pQCD) corrections,
while for the HT coefficients, possible pQCD effects have been neglected.

The nuclear effects in the deuteron are accounted by \eq{eq:IA} with the off-shell correction governed by \eq{delf}.
The deuteron AV18 wave function is used~\cite{Wiringa:1994wb,Veerasamy:2011ak}.
The function $\delta f(x)$ is determined along with the proton PDFs and HTs in a fit to the
deuterium data listed in Table~\ref{tab:data} and the proton data from Table~II in Ref.~\cite{Alekhin:2017fpf}.
This function is parametrized as a polynomial:
\begin{equation}\label{eq:deltaf}
\delta f(x)=c_0 + c_1 x + c_2 x^2.
\end{equation}
Note that \eq{eq:IA} describes the nuclear corrections driven by the momentum distribution, the nuclear binding, and the off-shell effect, which dominate in the present analysis.
We also verified~\cite{Alekhin:2017fpf} that the other nuclear effects, such as the meson-exchange currents and the nuclear shadowing, are within experimental uncertainties, and for this reason, they are not considered in the present analysis.

\section{Results}\label{sec:results}
\begin{figure}[p]
\centering
\vspace{-8ex}
\includegraphics[width=0.65\textwidth]{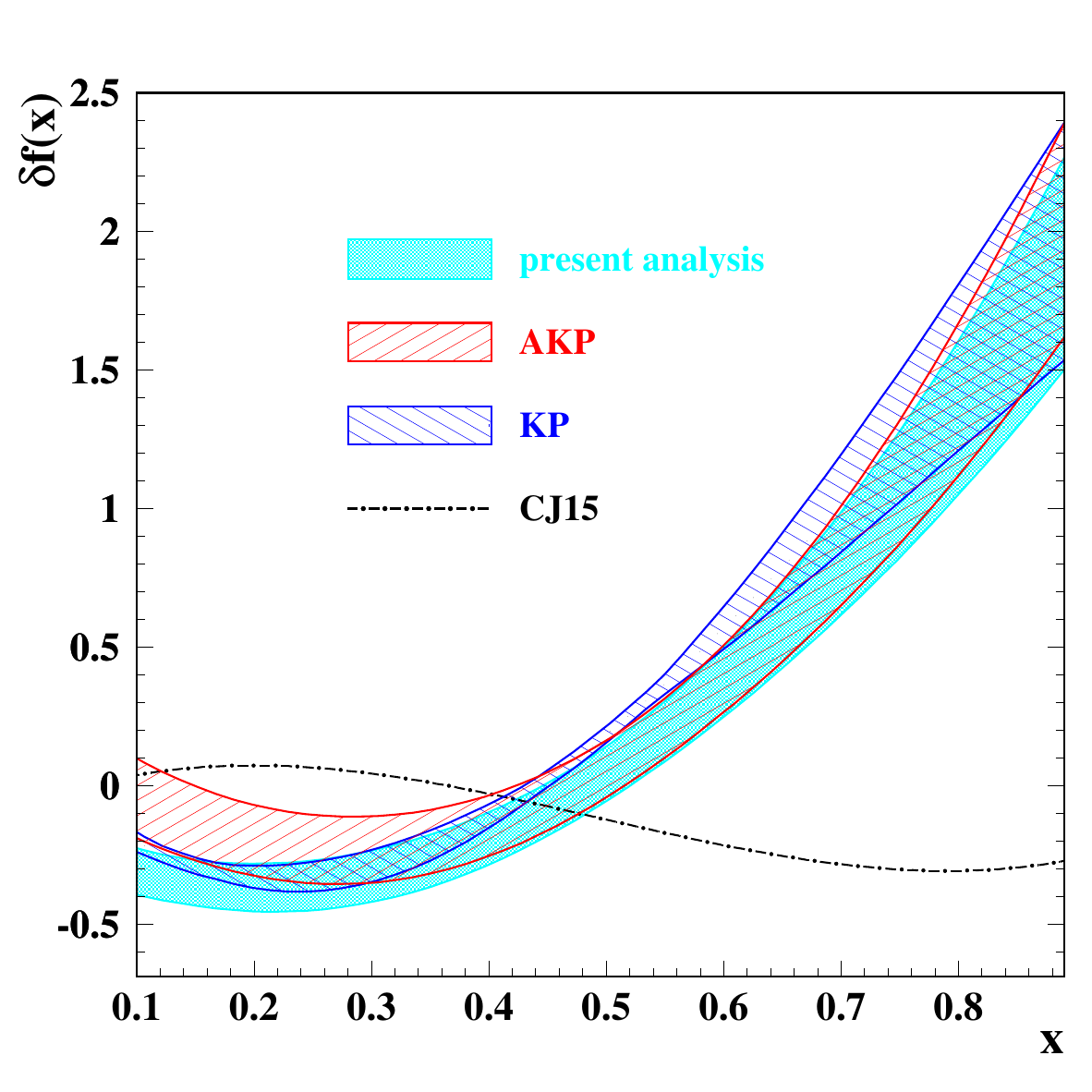}
\hfill
\caption{\label{fig:akpupd}
The $1\sigma$ uncertainty band for the off-shell function $\delta f$ as a function of
$x$ obtained in the present analysis (shaded cyan area) in comparison
with the results of the earlier AKP~\cite{Alekhin:2017fpf} (right-tilted hash area),
KP~\cite{Kulagin:2004ie} (left-tilted hash area), and  CJ15~\cite{Accardi:2016qay} (dash-dotted curve) fits.
}
\includegraphics[width=0.65\textwidth]{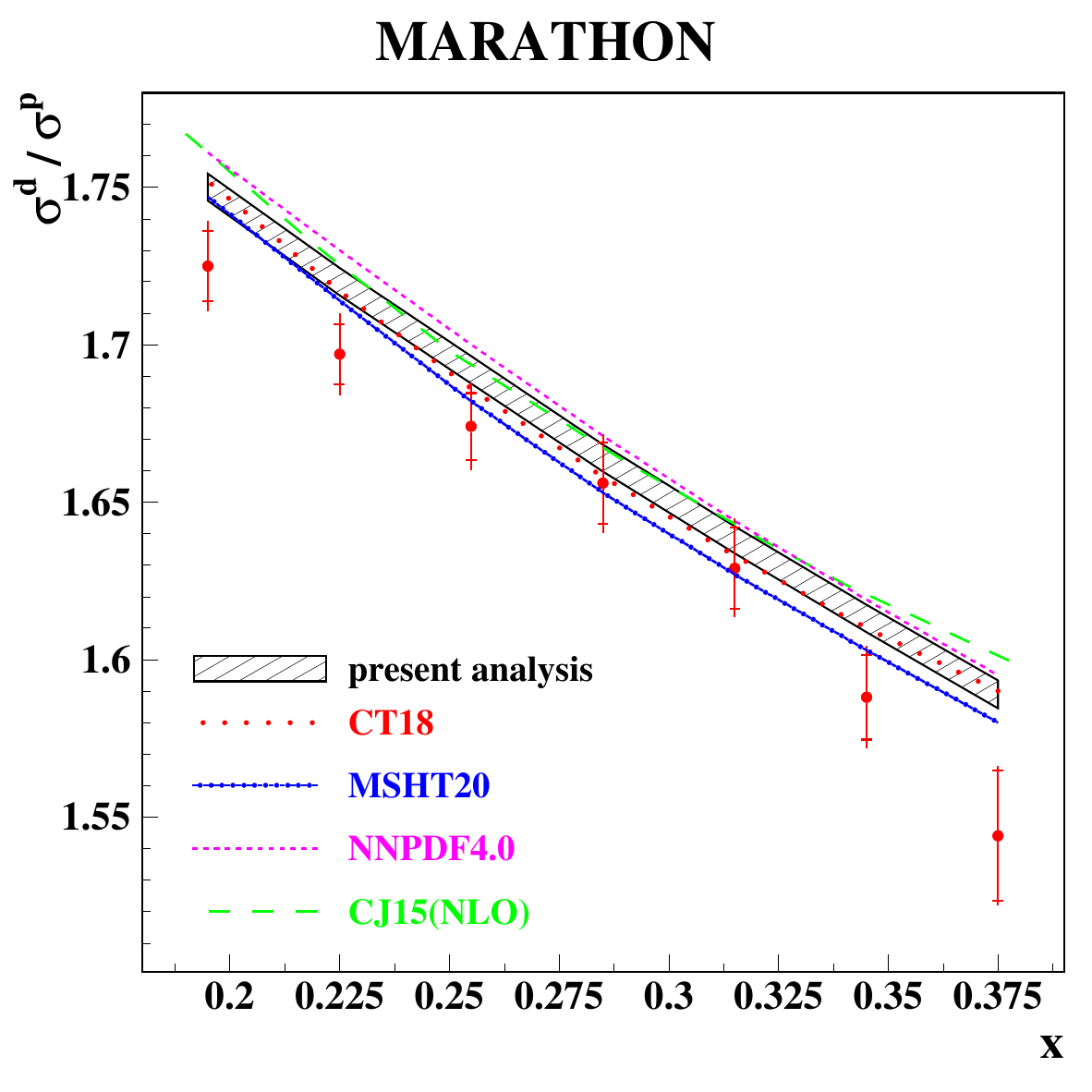}
\hfill
\caption{\label{fig:marathon}
Data on $\sigma^d/\sigma^p$ from \mara{} measurement~\cite{MARATHON:2021vqu} compared with corresponding $1\sigma$ band of present analysis (left-tilted area).
Also shown are the predictions of various global QCD analyses:
CT18~\cite{Hou:2019efy} (dots),
MSHT20~\cite{Cridge:2021qfd} (connected dots),
NNPDF4.0~\cite{NNPDF:2021njg} (short dashes), and
CJ15~\cite{Accardi:2016qay} (long dashes).}
\end{figure}

We simultaneously fit the parameters of the off-shell function $\delta f$ with those of the PDFs and HTs
in order to provide a consistent separation of the various contributions to the SFs by exploiting the broad $x$ and $Q^2$ coverage of available data.
The resulting data normalization factors and $\chi^2$ values corresponding to various deuterium datasets
are listed in Table~\ref{tab:data}, and the parameters of the $\delta f(x)$ function are
$c_0=-0.16\pm0.11$,
$c_1=-2.04\pm0.73$,
and $c_2=4.86\pm1.13$.
This function
is shown in Fig.~\ref{fig:akpupd}, together with the
results of other determinations of this quantity from
Refs.~\cite{Kulagin:2004ie,Alekhin:2017fpf,Accardi:2016qay}.
For all data points included in our fit we have $\chi^2/\text{d.o.f.}=4842/4044$.
The present results are in a good agreement with both our former global QCD analysis~\cite{Alekhin:2017fpf} and the analysis of Ref.~\cite{Kulagin:2004ie},
in which the function $\delta f(x)$ was determined
from a global fit to the data on the ratios $\sigma^A/\sigma^d$ for the
DIS cross sections off nuclear targets with the mass number $4\leq A\leq 208$
using the proton and neutron SFs of Ref.~\cite{Alekhin:2007fh}.
However, our results are in a strong disagreement with those of Ref.~\cite{Accardi:2016qay}.
Below we trace possible reasons of the discrepancy with Ref.~\cite{Accardi:2016qay}
by verifying the differences in the experimental datasets and in the underlying model.

In Fig.~\ref{fig:marathon} we compare our results with the recent measurement of $\sigma^d/\sigma^p$ by the \mara{} experiment~\cite{MARATHON:2021vqu}.
Also shown are the predictions from
the CT18~\cite{Hou:2019efy},
MSHT20~\cite{Cridge:2021qfd},
NNPDF4.0~\cite{NNPDF:2021njg}, and
CJ15~\cite{Accardi:2016qay} QCD analyses.
Note that the CJ15 analysis was performed to the NLO approximation, while all others were done to the NNLO one.
The CT18, MSHT20, and NNPDF4.0 structure functions are computed in the
3-flavour scheme using the code
\verb|OPENQCDRAD| (version 2.1)~\cite{openqcdrad:2016} combined with
the \verb|LHAPDF| (version 6) PDF grids~\cite{Buckley:2014ana,lhapdflib}
\verb|CT18NNLO|, \verb|MSHT20nnlo_nf3|, and \verb|NNPDF40_nnlo_pch_as_01180_nf_3|, respectively.
We use $F_2^d=F_2^p+F_2^n$ for CT18 and NNPDF4.0 in Fig.~\ref{fig:marathon}, as those analyses do not account for the deuteron correction.
Both MSHT20 and CJ15 account for the deuteron effect in their PDF fits.
For MSHT20, we take $F_2^d=R_d(F_2^p+F_2^n)$ with the correction factor $R_d$ obtained in the NNLO global QCD fit of Ref.~\cite{Cridge:2021qfd}.
For CJ15 we use their results obtained from Ref.~\cite{MARATHON:2021vqu}.
All the predictions are in agreement with \mara{} $\sigma^d/\sigma^p$ data within
uncertainties in the region about $x=0.3$ and somewhat overshoot the data for $x$
values about $0.2$ and $0.4$.%
\footnote{Note that in Fig.~\ref{fig:marathon} our predictions are for the ratio of cross sections $\sigma^d/\sigma^p$,
while for the other groups we compute the ratio $F_2^d/F_2^p$.
The relation $\sigma^d/\sigma^p=F_2^d/F_2^p$ is justified by observation that the deuteron and the nucleon have equal $R=\sigma_L/\sigma_T$ within experimental uncertainties~\cite{Arneodo:1996kd}.}

The quality of our fit for the newly added
\bonus{} data on $F_2^n/F_2^d$~\cite{Tkachenko:2014byy},
the cross-section measurements from JLab-E00-116 experiment~\cite{Malace:2009kw},
NMC data on $F_2^d/F_2^p$~\cite{Arneodo:1996kd}, and the HERMES data on $\sigma^d/\sigma^p$~\cite{Airapetian:2011nu}
are illustrated in detail in Figs.~\ref{fig:bonus} to \ref{fig:hermes} in Appendix~\ref{sec:pulls}. In general, we observe a good agreement of the fit and the data with no regular and/or statistically significant trend in the pulls.

\section{Systematic Studies}\label{sec:unc}
\begin{figure}[htb]
\vspace{-4ex}
\centering
\includegraphics[width=1.0\textwidth]{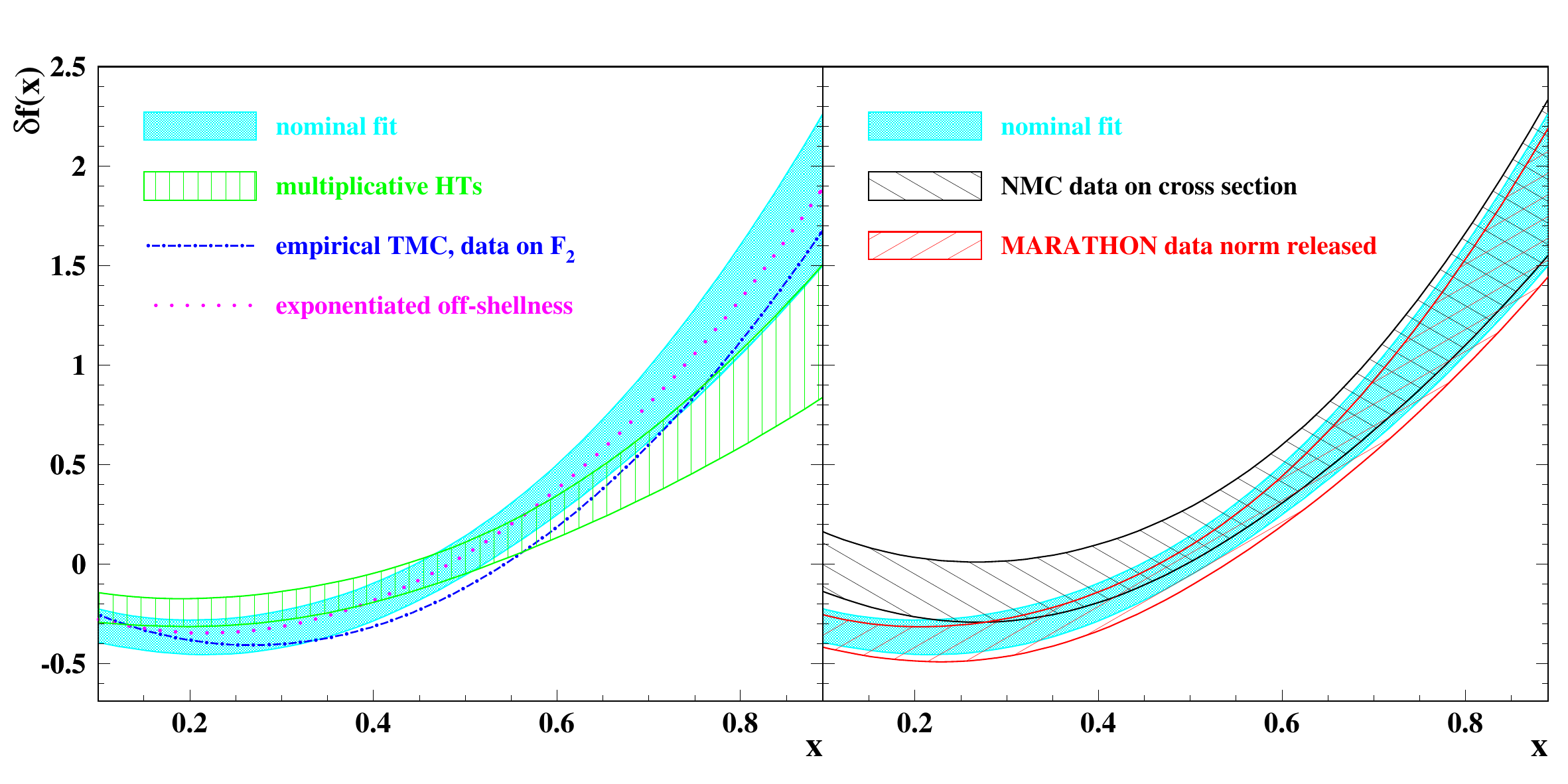}
\hfill
\vspace{-3ex}
\caption{\label{fig:akpunc}
The $1\sigma$ uncertainty band on $\delta f$ from our nominal fit
same as in Fig.~\ref{fig:akpupd} (shaded cyan area) compared with the results
of a modified fit framework.
Left panel:
higher twists (HTs) parametrized in a multiplicative ansatz by \eq{eq:htmult} (dashed curve);
using approximate TMC scheme by \eq{eq:tmc-approx} (see Eq.(61) of Ref.\cite{Schienbein:2007gr}) and the data on $F_2$ (dash-dotted curve);
exponential form of the off-shell correction by \eq{SF:OS2} (dotted curve).
Right panel:
using NMC cross-section data~\cite{Arneodo:1996qe} instead of NMC $F_2^d/F_2^p$ data~\cite{Arneodo:1996kd} in the nominal fit (right-tilted hash area);
released normalization of \mara{} $\sigma^d/\sigma^p$ data (left-tilted hash area).
}
\end{figure}

We verify the stability of our results on $\delta f$ by performing a number of fits with
a modified ansatz.
Some of these modifications, like the parametrization of higher-twist terms
(additive vs multiplicative form)
and the off-shell correction [a linear dependence on $v$ in \eq{SF:OS} vs exponentiated form of \eq{SF:OS2}],
reflect uncertainties in the theory framework of the fit.
The other modifications are motivated by other studies in the field and aimed to facilitate a comparison with those studies.
The impact of the modifications considered is summarized in Fig.~\ref{fig:akpunc}, and their detailed description is given below in this section.
Note that for all considered modifications of the fit, the data normalization factors are kept at the values of Table~\ref{tab:data}, thus allowing us to avoid the interplay with the data shift.

\subsection {NMC data choice}
\label{sec:unc:nmc}

Our fit includes the NMC data on the ratio $F_2^d/F_2^p$~\cite{Arneodo:1996kd}. These data are derived from the cross-section data assuming the same ratio $R=\sigma_L/\sigma_T$ for the proton and deuteron, which was verified experimentally by the NMC  with a good accuracy.
Alternatively, in our former analysis~\cite{Alekhin:2017fpf}, the cross-section data for the deuterium target~\cite{Arneodo:1996qe} have been employed instead. In order to verify the impact of the particular NMC data choice, we perform a variant of our nominal fit with the NMC data on $F_2^d/F_2^p$ replaced with the cross-section data on the deuterium target~\cite{Arneodo:1996qe}.
This fit results in $\chi^2/\text{d.o.f.}=4693/3988$.
As can be seen in Fig.~\ref{fig:akpunc}, the difference in the function $\delta f$ obtained in these two versions of the fit is significant only for $x\lesssim 0.4$. In this region, the uncertainty in the value of $\delta f$ extracted from the deuterium cross-section data is somewhat larger, due to less statistical significance of this sample.
Meanwhile, the error bands for the two determinations almost overlap with each other and with the
determination based on the heavy-nuclear  data~\cite{Kulagin:2004ie} (see also Fig.~\ref{fig:akpupd}).
Note also that $\delta f(x)$ obtained in the fit with the NMC deuterium cross-section data
is almost identical to our earlier result~\cite{Alekhin:2017fpf}.

\subsection{MARATHON data normalization}
\label{sec:unc:mara}

The normalization of the ratio $\sigma^d/\sigma^p$ in the \mara{} experiment is determined experimentally with a very high accuracy of 0.55\%. Nonetheless, the \mara{} data go somewhat lower than the other samples used in our analysis, cf. Fig.~\ref{fig:dp14x}. To quantify
this tension, we perform a variant of fit with the normalization of the \mara{} $\sigma^d/\sigma^p$ data released and adjusted
simultaneously with other fit parameters. The normalization factor of $1.014(4)$ obtained in this way is at about  $2\sigma$ off the nominal value $1.0000(55)$. However, this  renormalization of the \mara{} $\sigma^d/\sigma^p$ ratio has a negligible impact on the
value of $\delta f$ extracted from the data, and the corresponding change in its value is well within $1\sigma$ uncertainty band.
For this fit, we have $\chi^2/\text{d.o.f.}=4834/4044$.

\subsection{Higher-twist correction}
\label{sec:unc:ht}

\begin{figure}[htb]
\centering
\includegraphics[width=0.5\textwidth]{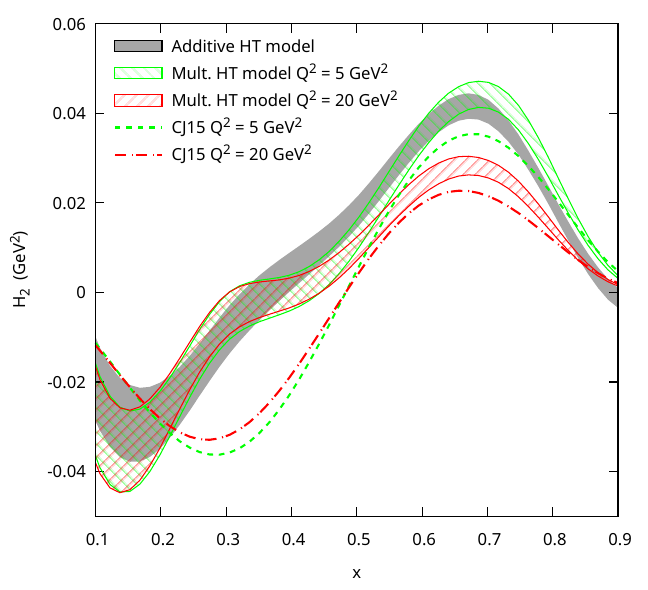}%
\includegraphics[width=0.5\textwidth]{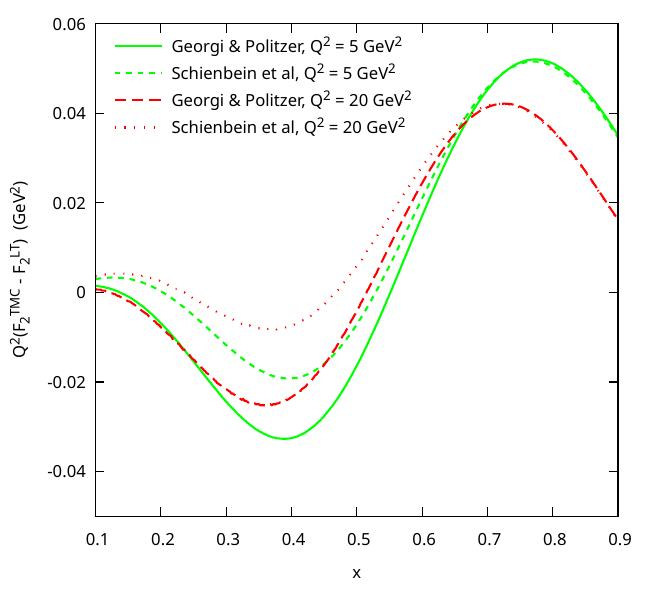}
\caption{%
\label{fig:httm}
Left panel:
The $1\sigma$ bands of twist-4 correction to the proton $F_2$ using the additive HT model [\eq{eq:sfdis}, gray band] and multiplicative HT model [\eq{eq:htmult}].
The multiplicative HT is shown for $Q^2=5$ (right-tilted band) and $20\gevsq$ (left-tilted band) in order to illustrate their residual $Q^2$ dependence.
Also shown are the corresponding CJ best fit results~\cite{Accardi:2016qay} (dashed and dashed-dotted curves).
Right panel:
Target mass correction for the proton $F_2$ in terms of $Q^2[F_2^\text{TMC}(x,Q^2)-F_2^\text{LT}(x,Q^2)]$
suitable for comparison with HT in the left panel.
Georgi-Politzer TMC~\cite{Georgi:1976ve} is shown for $Q^2=5$ and $20\gevsq$ (solid and long-dashed curves).
The short-dashed and dotted lines show an empirical TMC by \eq{eq:tmc-approx},
which was used in analysis~\cite{Accardi:2016qay}.
}
\end{figure}

In our analysis we compute the structure functions following \eq{eq:sfdis}
with an additive model of the higher-twist (HT) terms motivated by the OPE.
However, a multiplicative HT model is often used in the literature (see, e.g., Refs.~\cite{Virchaux:1991jc,Accardi:2016qay,Cocuzza:2021rfn}):
\begin{equation}\label{eq:htmult}
F_{i}(x,Q^2,p^2) = F_{i}^{\text{TMC}}(x,Q^2,p^2)
+ F_{i}^{\text{LT}}(x,Q^2) h_{i}(x)/ Q^{2},
\end{equation}
where $i=2,T$. To compare the additive and multiplicative HT models, one should confront the coefficients $H_{i}$  in \eq{eq:sfdis} with the corresponding product $F_{i}^\text{LT}h_{i}$ in \eq{eq:htmult}.
These terms have a different $Q^2$ dependence driven by assumptions
about anomalous dimensions of the HT operators:
For the additive form, they are neglected, and for the multiplicative one, they are similar to the leading-twist case.
This difference is  important
at large $x$, where the leading-twist evolution and the TMC are most significant (for illustration, see  Fig.~\ref{fig:httm}).
The same trend  appears in the determination of the off-shell function,
which is sensitive to the assumed HT model at
$x\gtrsim 0.5$, although the shape of $\delta f(x)$ (left panel of Fig.~\ref{fig:httm})
does not change essentially under the HT model variation.
\updtxt{%
For the fit with multiplicative HT model $\chi^2/\text{d.o.f.}=4798/4044$.
}

\subsection{Target mass effect}
\label{sec:unc:tmc}

The target mass effects are taken into account in our analysis using the Georgi-Politzer formalism in the off-shell region [see \eqs{eq:TMC}].
In other global QCD fits, the TMC is either neglected or treated differently, assuming $p^2=M^2$.
For example, in the CJ15 analysis~\cite{Accardi:2016qay,Accardi:pc}
the TMC is accounted using an approximation to \eq{eq:TMC} (see Eq.(61) in  Ref.~\cite{Schienbein:2007gr}):
\begin{equation}\label{eq:tmc-approx}
F_2^{\text{TMC}}(x,Q^2) \approx
	\frac{(1+\gamma)^2}{4 \gamma^3} F_2^\text{LT}(\xi,Q^2)
	\left[ 1 +	\frac{3(\gamma-1)}{\gamma}(1-\xi)^2 \right].
\end{equation}
To verify the sensitivity of our results to a particular TMC treatment,
we perform a fit using \eq{eq:tmc-approx} instead of \eq{eq:TMC:2}.
A similar approximation for $F_T^{\text{TMC}}$ is not available in the formalism of Ref.~\cite{Schienbein:2007gr}.
For this reason, in our modified fit, we employ data on $F_2$, instead of cross-section data,
for the SLAC, CERN-BCDMS, CERN-NMC, and  JLab-E00-116 experiments.
The DESY-HERMES and MARATHON data on the cross-section ratio are treated using the relation
$\sigma^d/\sigma^p\approx F_2^d/F_2^p$ since $R=\sigma_L/\sigma_T$ is similar
for proton and deuteron at moderate $x$~\cite{Arneodo:1996kd}.
Finally, for the HERA cross-section data, we take $F_T$ according to \eq{eq:TMC:T}.
Such an approach does not cause a serious inconsistency due to the HERA data
being localized at small $x$, where the TMC is generally small.
For this variant of the fit, we have $\chi^2/\text{d.o.f.}=4831/4050$.
The best fit result on $\delta f(x)$ is shown by the dashed-dotted curve in Fig.~\ref{fig:akpunc}, which is
within the $1\sigma$ band of our nominal fit for almost all $x$ values but a region around $x=0.4$. In this region \eq{eq:tmc-approx} gives a rather poor approximation on \eq{eq:TMC:2},
as illustrated in Fig.~\ref{fig:httm} (right panel).
The maximal difference between the two implementations is observed in the range $0.2 \leq x \leq 0.5$. Note that in the same region we observe opposite deviations between the HT contributions (left panel of Fig.~\ref{fig:httm}) obtained by the CJ15 analysis and by our fit using the same multiplicative HT form and the TMC from \eq{eq:TMC}.

\subsection{Exponential model of off-shell correction}
\label{sec:unc:off}

In our present study, we compute the off-shell correction using \eq{SF:OS}.
As noted in Sec.~\ref{sec:pn}, at high nucleon momentum $|\bm p|\gtrsim M$ in
the nuclear convolution \eq{eq:IA} the off-shell structure function from \eq{SF:OS}
may be negative, thus signaling a violation of the linear approximation in $v$.
To verify the relevance of \eq{SF:OS}, we performed a fit using the exponential model
of off-shell correction from \eq{SF:OS2}, which gives a positive SF at any value of $v$.
The resulting function $\delta f(x)$, shown in Fig.~\ref{fig:akpunc},
is identical to our nominal fit result within the fit uncertainties
and $\chi^2/\text{d.o.f.}=4847/4044$ for this variant of the fit.

\section{Discussion}
\label{sec:discus}

In Sec.~\ref{sec:results} and \ref{sec:unc}, we discussed determination of the quantity $\delta f(x)$
from a global QCD analysis of the most recent DIS data off hydrogen and deuterium combined with the ones on $W$- and $Z$-boson production at hadron colliders.
We reiterate that $\delta f(x)$ describes the modification of the nucleon PDFs in the off-shell region for bound nucleons and, as such,
it is expected to be a universal quantity independent of the nucleus considered.
This quantity has a considerable impact on the
nuclear corrections obtained within the nuclear convolution approach
and is required to describe available nuclear DIS~\cite{Kulagin:2004ie,Kulagin:2010gd} and
Drell-Yan data~\cite{Kulagin:2014vsa,Ru:2016wfx}.
The results from our current analysis are in good agreement with our previous analysis~\cite{Alekhin:2017fpf},
as well as with the study of nuclear ratios of DIS cross sections for nuclear targets with
nuclear number $A\geq 3$~\cite{Kulagin:2004ie,Kulagin:2010gd} (Fig.~\ref{fig:akpupd}).
However, our results disagree with the ones of Ref.~\cite{Accardi:2016qay}.

In our previous analysis~\cite{Alekhin:2017fpf}, we evaluated the uncertainties associated with the modeling of the deuteron wave function
and with the use of different datasets. In Sec.~\ref{sec:unc}, we performed additional systematic studies on both the input ansatz and the datasets
in order to verify our results and to further investigate the observed discrepancies with the analysis of Ref.~\cite{Accardi:2016qay}.
In all cases, our results on $\delta f$ are stable against the modifications of the fit considered, and the corresponding variations
are generally consistent with the quoted uncertainties.
We find that while $\delta f$ has some sensitivity to the implementation of the HT corrections (i.e., additive vs multiplicative)
at large Bjorken $x$, its shape is essentially unchanged (Fig.~\ref{fig:akpunc}).
In general, the systematic uncertainties related to the use of different deuterium datasets,
in particular NMC cross-section data vs NMC $F_2^d/F_2^p$ ones,
are comparable to the ones related to the input model assumptions.

\begin{figure}[p]
\centering
\includegraphics[width=1.0\textwidth]{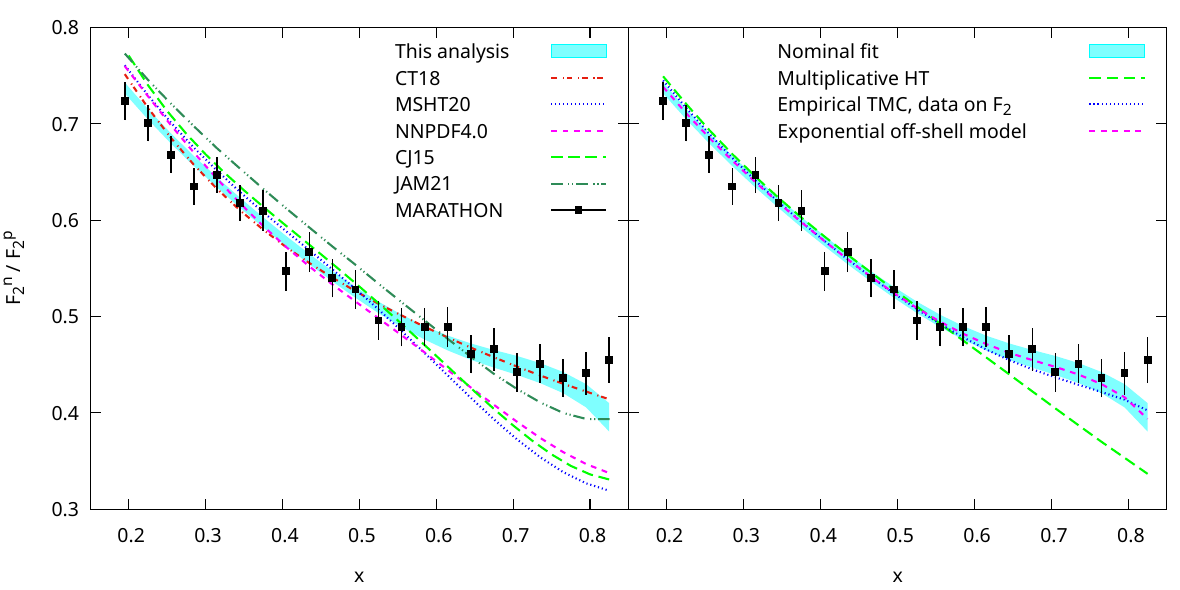}
\hfill
\caption{\label{fig:f2nf2p}
Data from the \mara{} measurement of $F_2^n/F_2^p$~\cite{MARATHON:2021vqu} compared with the predictions of the present analysis.
In the left panel, shown are the predictions from recent PDF analyses: CT18~\cite{Hou:2019efy} (dash-dotted line), MSHT20~\cite{Cridge:2021qfd} (dotted line), NNPDF4.0~\cite{NNPDF:2021njg} (short dashes), CJ15~\cite{Accardi:2016qay} (long dashes),
and JAM21~\cite{Cocuzza:2021rfn} (dash-double-dot).
In the right panel, shown are our predictions based on modified fit results discussed in Sec.~\ref{sec:unc}.
}
\includegraphics[width=1.0\textwidth]{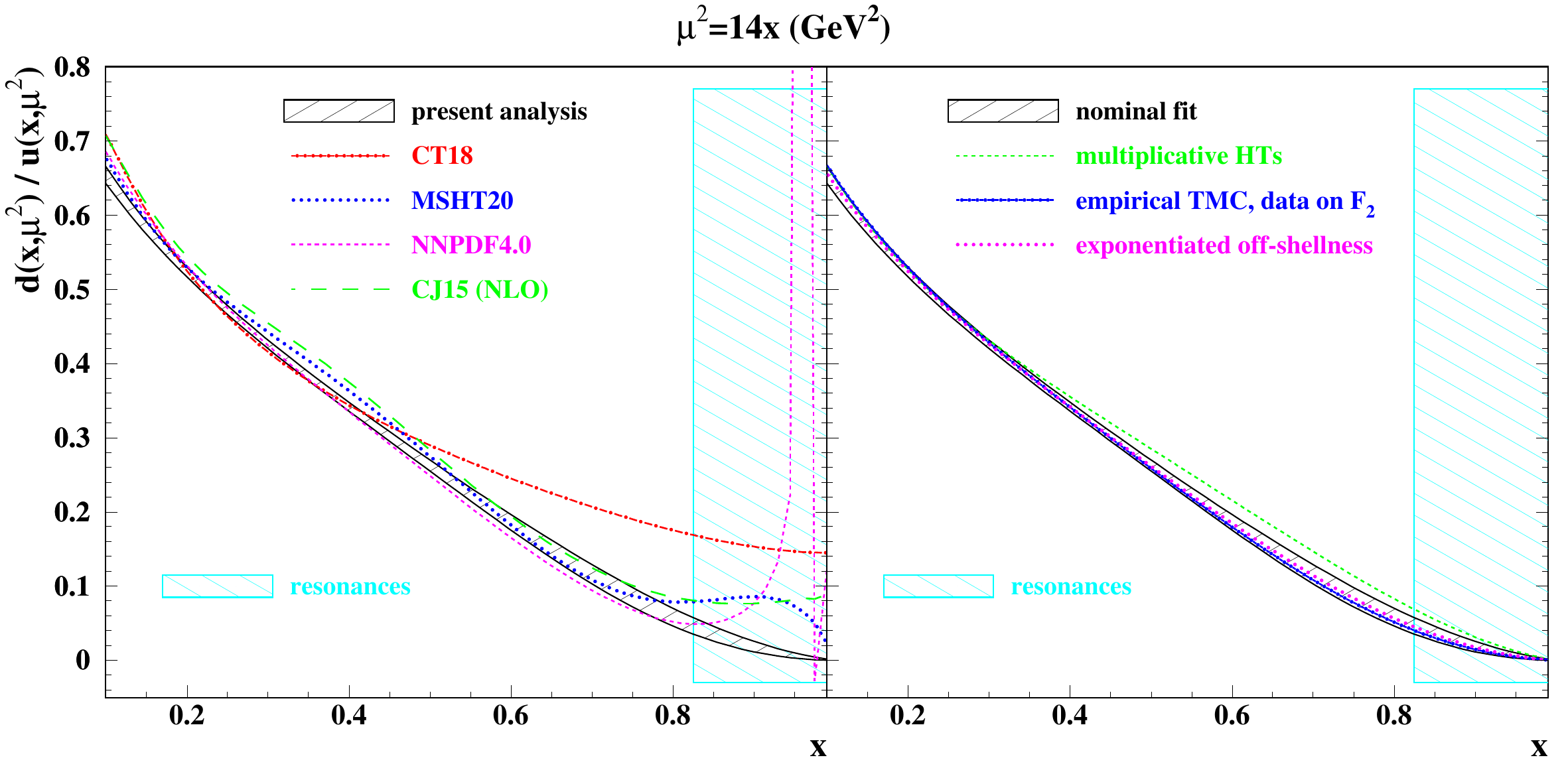}
\hfill
\caption{\label{fig:du}
The $1\sigma$ band of the PDF ratio $d/u$ vs. $x$ computed at variable scale $\mu^2=14x\ (\mathrm{GeV}^2)$ (left-tilted area).
Also shown are the results from PDF analyses, same as in Fig.~\ref{fig:f2nf2p} (left panel).
The right-tilted (yellow) area shows the nucleon resonance region of $W\lesssim 2\gev$.
All shown analyses but Ref.~\cite{Accardi:2016qay} are to NNLO pQCD order.
Right panel shows our results from modified fits, same notations as in Fig.~\ref{fig:f2nf2p}.}
\vfill
\end{figure}

We further verify our results by comparing our predictions on the
ratio $R_{np}=F_2^n/F_2^p$ with the recent \mara{} data~\cite{MARATHON:2021vqu}, which were not included in our fits.
The calculations are performed for the \mara{} kinematics,
which is roughly consistent with $Q^2=14x\ (\mathrm{GeV}^2)$, and are shown in Fig.~\ref{fig:f2nf2p}.
Our independent predictions for $R_{np}$ are in excellent agreement with the \mara{} measurement over the entire $x$ range available.

In Fig.~\ref{fig:f2nf2p} (left panel), we show the predictions on $R_{np}$ obtained from the QCD analyses CJ15, MSHT20, NNPDF4.0, CT18, and JAM21.
The structure functions are computed as described in Sec.~\ref{sec:results}.
The calculations include the target mass corrections of Ref.~\cite{Georgi:1976ve}.
The CJ15 prediction is obtained from Fig.~3 of Ref.~\cite{MARATHON:2021vqu}.
At large Bjorken $x$ ($x>0.6$), significant differences are observed.
While the predictions from CJ15, MSHT20, and NNPDF4.0 are
consistent with each other, they differ substantially from CT18 and our results.
Note that different assumptions on the HT contributions are used for the various calculations:
additive HT for our results, multiplicative HT for CJ15, and no HT (only LT term) for the others.
To verify the impact of the $d$-quark PDF on such differences, we compare the corresponding predictions for the
$R_{du}=d/u$ ratio for the \mara{} kinematics using the PDFs from the LHAPDF library~\cite{lhapdflib}.%
\footnote{%
As noted in Ref.~\cite{Accardi:2021ysh}, the nuclear correction at large $x$
also affects the valence $d$ quark PDF at small values of $x\approx 0.03$,
because of fermion number conservation.
While this effect is relevant for the electroweak studies at LHC,
we leave its analysis for
future, as in the present work we focus on the region relevant for the \mara{} measurement.
}
Figure~\ref{fig:du} shows our $1\sigma$ uncertainty band together with the central values obtained for the other analyses.
The uncertainties on these latter are relatively large for $x>0.6$ due to the use of tight $W$ cuts, which effectively
exclude high-$x$ DIS data from the QCD analysis.%
\footnote{As an illustration, the CT18 uncertainty on $R_{du}$ is about 100\% as can be seen in Fig.~9 of Ref.~\cite{Hou:2019efy}.}
Figure~\ref{fig:f2nf2p} indicates that the CJ15-, MSHT20-, and NNPDF4.0-based predictions are in clear disagreement
with the \mara{} $R_{np}$ data.

The effect of the variations of the model assumptions discussed in Sec.~\ref{sec:unc} on our predictions for $R_{np}$
is also illustrated in Fig.~\ref{fig:f2nf2p} (right panel). Although in most cases the results are consistent with the nominal one within
the quoted uncertainties, a significant deviation is apparent for the variant with multiplicative HT contributions from
\eq{eq:htmult}. In this latter case our predictions for $R_{np}$  appear to be closer to the CJ15-, MSHT20- and NNPDF4.0-based calculations.
However, the corresponding $R_{du}$ ratio shown in Fig.~\ref{fig:du} (right panel) is consistent with our nominal fit.
These observations indicate that the \mara{} data are sensitive to the HT contributions in the region $x>0.6$.

The comparison with the CT18-based predictions is instructive, as they provide a good LT description of the
\mara{} data on $R_{np}$ (Fig.~\ref{fig:f2nf2p}) without any HT contributions. This agreement is explained by
a much larger value of the $R_{du}$ ratio at $x>0.6$
compared to the other QCD analyses,
as illustrated in Fig.~\ref{fig:du} (left panel).%
\footnote{\updtxt{%
We note that the CT18 analysis utilizes neutrino data from heavy nuclear targets.
The value of $R_{du}$ is significantly reduced at large $x$ if neutrino data are removed from the analysis~\cite{Accardi:2021ysh}.
}}
By contrast, we obtain a good description of \mara{} data on $R_{np}$ with $R_{du}\to0$ as $x\to1$ and
a sizable HT contribution, which is maximal at $x\sim 0.7$ (Fig.~\ref{fig:httm}).
In general, $W^\pm$ boson production and the corresponding lepton asymmetries from D0 and LHCb data
at high rapidity could help to clarify the differences observed on the $d$-quark distribution at large Bjorken $x$~\cite{Alekhin:2015cza,Alekhin:2018dbs}.
However, the calculation of the cross-section for $W$-boson production in the NNLO pQCD approximation
suffers from uncertainties in the available numerical codes~\cite{Alekhin:2021xcu}.

As discussed in Sec.~\ref{sec:pn}, we assume isoscalar HT terms in the additive HT model ($H^p=H^n$).
We also assume  $h^p=h^n$ for the multiplicative HT model.
However, in the latter case the overall HT correction is different for
protons and neutrons since $h$ is multiplied by the corresponding LT structure functions.
Note that for the multiplicative HT model the contribution from
$h$ cancels out
in the ratio $R_{np}$, which has a similar behavior
as the corresponding LT approximation.
Conversely, in the case of the additive HT model used
in our nominal fit
the ratio $R_{np}$ receives a finite HT correction.
The \mara{} $R_{np}$ data seem to prefer a common additive HT contribution (Fig.~\ref{fig:f2nf2p}) for
both the neutron and the proton.
Although the disagreement observed for $x>0.6$ with the multiplicative HT form
may be mitigated by the introduction of an explicit isospin dependence in the $h$ terms, such an effect
could result in observable deviations for other DIS data sensitive to isospin effects.

A recent paper~\cite{Cocuzza:2021rfn} reports the results of a global QCD analysis (JAM21) including
\mara{} data on the cross-section ratios $\sigma^d/\sigma^p$ and $\sigma^\text{3H}/\sigma^\text{3He}$ for the three-body nuclei,
as well as the previous measurement of $\sigma^\text{3He}/\sigma^d$ from E03-103 at JLab~\cite{Seely:2009gt}.
The study includes multiplicative HT corrections and
a calculation of nuclear effects based on the
convolution approach supplemented by off-shell corrections. However, the treatment of these latter corrections
is rather different with respect to our implementation. In the JAM21 analysis, the off-shell correction
depends on both the specific nucleus considered and on the isospin of the target nucleon (different for
protons and neutrons), resulting in multiple functions that are extracted from data.
In particular, the need of an explicit isovector contribution in the off-shell functions
is advocated to describe the \mara{} data.
\updtxt{%
Our analysis indicates that this result may be affected by the assumption of multiplicative HT terms.
}
The result of the JAM21 fit on $R_{np}$ is shown in the left panel of Fig.~\ref{fig:f2nf2p} and appears to be in disagreement with the \mara{} \updtxt{$F_2^n/F_2^p$ data}~\cite{MARATHON:2021vqu}.

The determination of $\delta f$ described in this paper is based on deuterium DIS data and is therefore only sensitive to the
isoscalar combination $F_2^p+F_2^n$.
Our results are in a good agreement with the study of the ratios of nuclear DIS cross sections $\sigma^A/\sigma^d$ with nuclear number
$A \geq 3$, in which the nuclear EMC effect  was successfully described in terms of a nuclear convolution approach with
\updtxt{a universal off-shell function $\delta f(x)$ independent of the nucleus.}
Although the model~\cite{Kulagin:2004ie} could naturally incorporate an isospin dependence into the off-shell correction,
the good agreement with data on nonisoscalar nuclei obtained using the
\updtxt{same off-shell function $\delta f(x)$ for the proton and neutron~\cite{Kulagin:2004ie,Kulagin:2010gd}}
seems to indicate that potential \updtxt{isospin dependence of $\delta f$ is small.}
Dedicated studies of nuclear effects using DIS data from mirror nuclei $^3$H and $^3$He~\cite{MARATHON:2021vqu} and
upcoming DIS data on proton and deuterium from JLab12~\cite{E1210002}
could provide new insights on the origin of modification of parton structure in bound nucleons,
as well as improved constraints on nucleon $d$-quark distribution at large $x$ and on the isospin dependence of HT corrections.
A more comprehensive study of these effects would require future data from high-energy processes, which can provide a
flavor selection like the hadronic Drell-Yan reaction or DIS using both the electron and (anti)neutrino charged-current (CC) process. To this end, the availability of precision measurements at future electron-ion collider~\cite{AbdulKhalek:2021gbh} and of both neutrino and antineutrino CC interactions off hydrogen and various nuclear targets~\cite{Petti:2019asx,Duyang:2018lpe} at the long-baseline neutrino facility could provide valuable insights.

\acknowledgments
\noindent
We would like to thank
A.~Accardi for useful discussions and communications on the CJ15 results,
G.~G.~Petratos for useful communications on the \mara{} results,
M.~V.~Garzelli and S.-O.~Moch for useful discussions and a careful reading of the manuscript, and Fr.~Kok for the help with manuscript preparation.
The work of S.A. is supported by the DFG grants MO 1801/5-1, KN 365/14-1.
R.P. is supported by Grant No. DE-SC0010073 from the Department of Energy, USA.

\newpage
\appendix
\section{Comparisons of the fit results with data}
\label{sec:pulls}

\begin{figure}[htb!]
\centering
\includegraphics[width=1.0\textwidth]{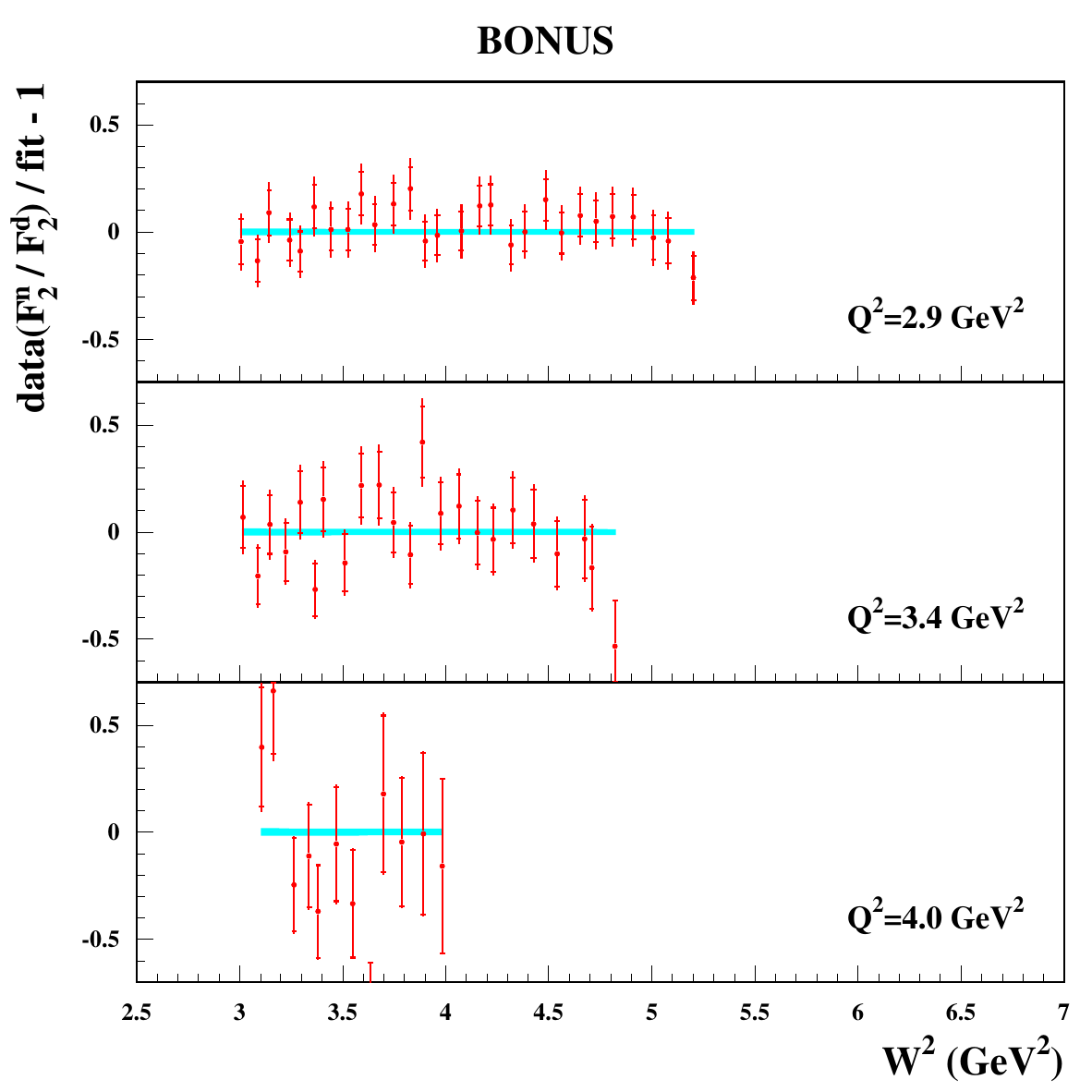}
\hfill
\caption{\label{fig:bonus}
The pulls of the present analysis for the \bonus{} data on the ratio $F_2^n/F_2^d$~\cite{Tkachenko:2014byy} vs $W^2$ displayed in the panels of $Q^2$ bins.
The inner error bars reflect the statistical and uncorrelated systematic uncertainties of data, while the outer bars are the total experimental error.
The shaded region shows $\pm1\sigma$ uncertainty of the fit.
}
\end{figure}
\begin{figure}[p]
\centering
\includegraphics[width=1.0\textwidth]{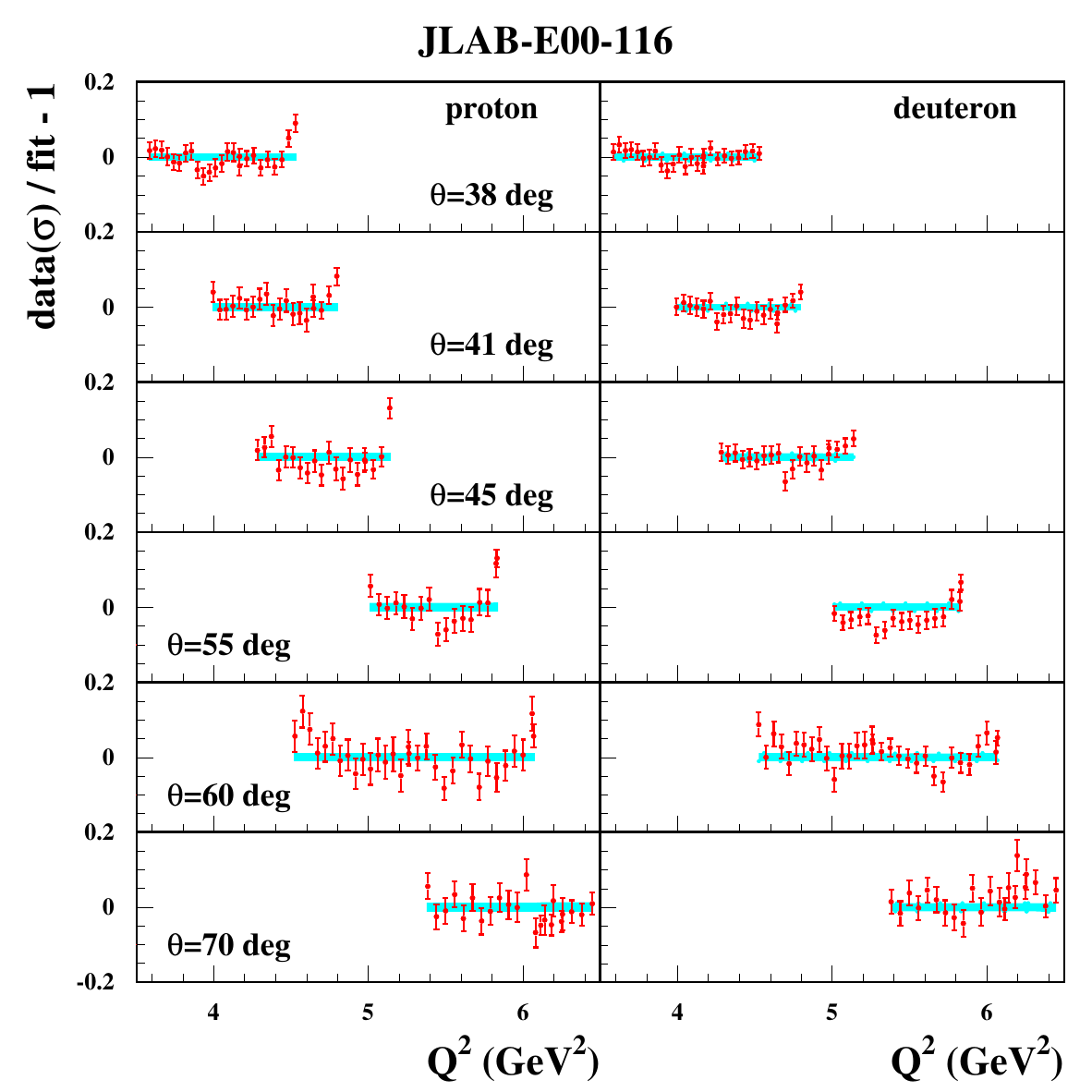}
\hfill
\caption{\label{fig:e116}
The pulls of the JLab-E00-116 data~\cite{Malace:2009kw} on $\frac{\ud^2\sigma}{\ud E^\prime \ud\Omega}$ for the proton (left panels) and the deuteron (right panels) targets vs $Q^2$. The panels correspond to the bins of the electron scattering angle $\theta$. Notations are similar to those in Fig.~\ref{fig:bonus}.
}
\end{figure}
\begin{figure}[p]
\centering
\includegraphics[width=1.0\textwidth]{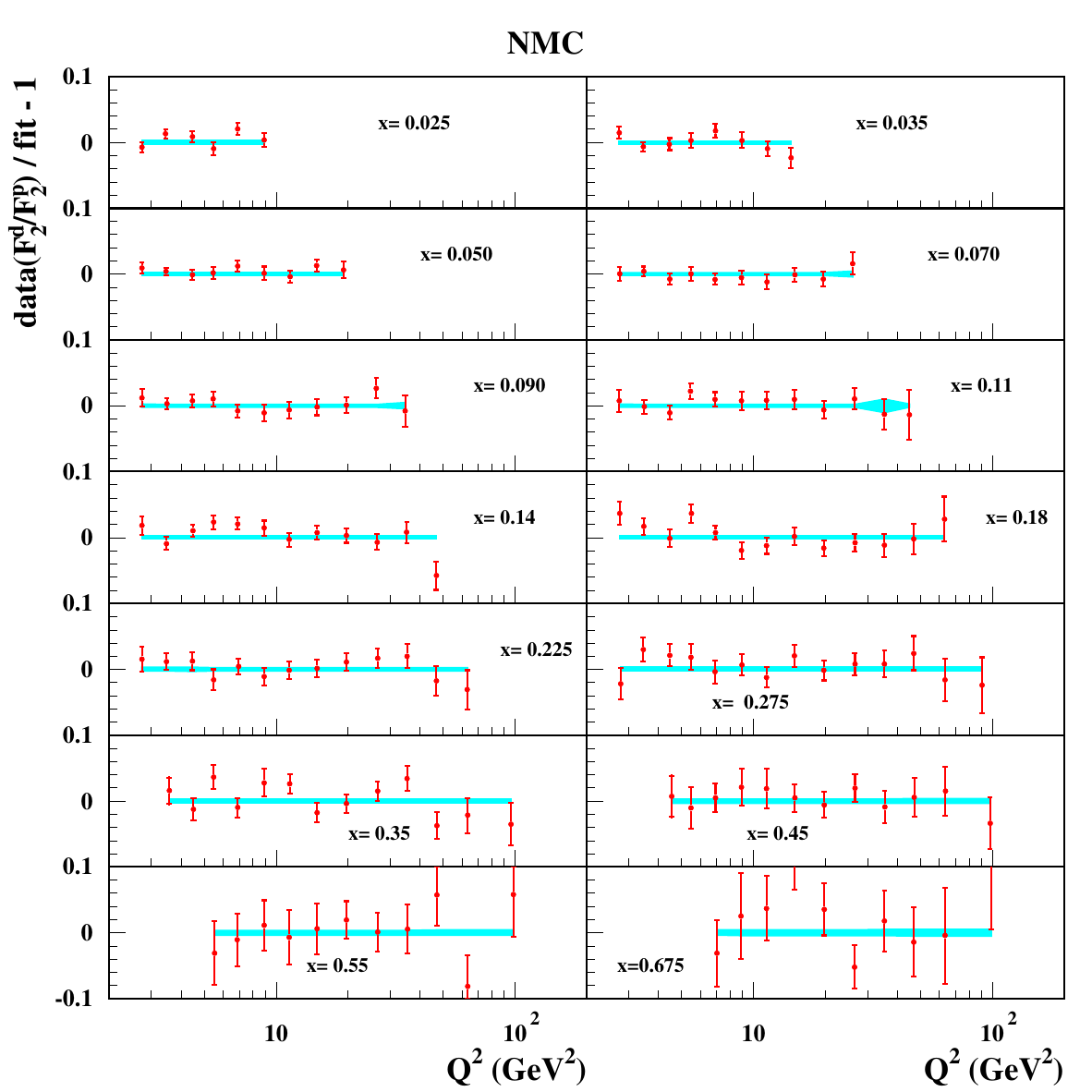}
\hfill
\caption{\label{fig:nmc}
The pulls of the NMC data~\cite{Arneodo:1996kd} on $F_2^d/F_2^p$ vs $Q^2$. The panels correspond to the bins of $x$. Notations are similar to those in Fig.~\ref{fig:bonus}.
}
\end{figure}

\begin{figure}[p]
\centering
\includegraphics[width=1.0\textwidth]{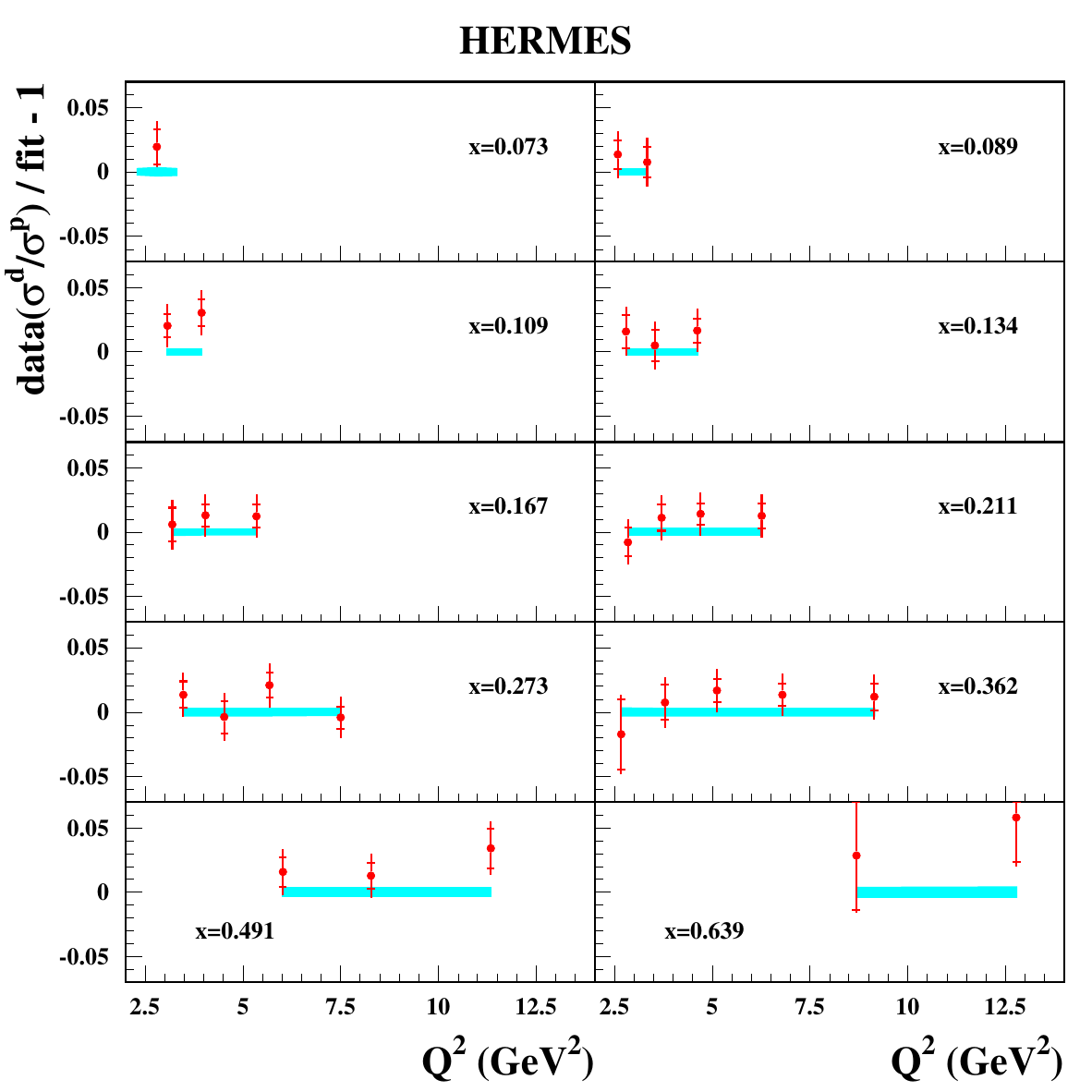}
\hfill
\caption{\label{fig:hermes}
The pulls of the DESY-HERMES data~\cite{Airapetian:2011nu} on $\sigma^d/\sigma^p$ vs $Q^2$. The panels correspond to the bins of $x$. Notations are similar to those in Fig.~\ref{fig:nmc}.
}
\end{figure}

\newpage
\section{Phase space integration in the nuclear convolution}
\label{sec:psi}

Here, we discuss in more detail the integration in \eq{eq:IA}
for general kinematics of $x$ and $Q^2$.
Recall that integration region is constrained  by the condition,
\begin{equation}\label{eq:w2ineq1}
	W^2 \ge W_\text{th}^2,
\end{equation}
where $W^2=(p+q)^2$ and $p$ is the four-momentum of the bound nucleon, and $W_\text{th}$ is the threshold mass.
The pion production threshold corresponds to $W_\text{th}=M+m_\pi$, and by setting $W_\text{th}=M$, we also include the elastic channel.
Because of the energy-momentum conservation, $p=p_d-p_S$, where $p_d$ is the deuteron four-momentum,
and $p_S=(\sqrt{m_S^2+\bm p^2}, -\bm p)$, the four-momentum of the spectator system with the mass $m_S$.%
\footnote{%
For scattering off the deuteron, $m_S=M$.
In case of scattering off a nucleus of $A$ nucleons,
$m_S$ is the mass of the residual nucleus of $A-1$ nucleons.}
In Appendix~A of Ref.\cite{Kulagin:2004ie}, the nuclear convolution integral with the constraint (\ref{eq:w2ineq1})
was considered for nonrelativistic spectator assuming $\sqrt{m_S^2+\bm p^2}=m_S+\bm p^2/(2m_S)$.
This approximation makes sense as the deuteron is a weakly bound system and most of the momentum distribution is in the  nonrelativistic region.
However, the high-momentum part with $|\bm p|$ of order of a few hundred MeV requires a relativistic analysis.
Here, we discuss the fully relativistic case of spectator kinematics that would allow us to better describe the contribution from the high-momentum region of the spectator.

In terms of the four-vectors $p_d$, $q$, and $p_S$, we can write \eq{eq:w2ineq1} as follows:
\begin{align}\label{eq:w2ineq2}
	(p_d+q)^2 + m_S^2 - 2(p_d+q)\cdot p_S \ge W_\text{th}^2.
\end{align}
In order to facilitate the discussion of \eq{eq:w2ineq2}, we use the following notations
\begin{subequations}\label{eq:nota}
\begin{align}
	S &= (p_d+q)^2 = M_d^2+Q^2\left(1/x_d - 1\right),
\\
	E & = M_d+q_0 = \sqrt{S+\bm{q}^2},
\\
	\alpha &= (S+m_S^2-W_\text{th}^2)/(2E m_S),
\\
	\beta &= |\bm{q}|/E,
\end{align}
\end{subequations}
where $S$ and $E$ are, respectively, the invariant mass squared and the energy of the virtual photon--deuteron system, and
$x_d=Q^2/(2 p_d\cdot q)$ is the natural Bjorken variable for the deuteron.
As it follows from the definitions in Eqs.(\ref{eq:nota}),  $\alpha>0$, and $0<\beta<1$ at any finite $Q^2$ value.
In the limit $Q^2\to\infty$, we have $\beta=1$ and $\alpha=(1-x_d)M_d/m_S$.
Using Eqs.(\ref{eq:nota}), we can write Eq.(\ref{eq:w2ineq2}) as follows:
\begin{equation}\label{ineq:D2}
	\alpha m_S - \sqrt{m_S^2+\bm{p}^2} + \beta p_z \ge 0.
\end{equation}
For completeness, we also give here \eq{ineq:D2} for nonrelativistic spectator kinematics
\begin{equation}\label{ineq:D2nr}
	2(\alpha-1) m_S^2 - \bm{p}^2 + 2\beta m_S p_z \ge 0.
\end{equation}
Below we discuss the solution to \Eqs{ineq:D2}{ineq:D2nr} in terms of both
the spherical coordinates and the $(p_z,\bm p_\perp)$ basis for the momentum $\bm p$.

\subsection{Convolution integral using spherical coordinates}
\label{sec:conv2:spherical}

We consider \Eqs{ineq:D2}{ineq:D2nr}  in spherical coordinates,
in which $p_z=|\bm p|\cos\theta$ with $\theta$ the zenith angle. 
Both \Eqs{ineq:D2}{ineq:D2nr} have two nodes, for which we will use the notation $p_\pm(\cos\theta)$.
For \eq{ineq:D2}, we have
\begin{equation}\label{p:bounds:D}
	p_\pm(\cos\theta) = m_S
	\frac{\alpha \beta \cos\theta \pm \sqrt{\beta^2\cos^2\theta + \alpha^2 - 1}}
	{1-\beta^2 \cos^2\theta}  ,
\end{equation}
while for \eq{ineq:D2nr}, we have
\begin{equation}\label{p:bounds:D:nr}
	p_\pm(\cos\theta) = m_S
	\left(
	\beta \cos\theta \pm \sqrt{\beta^2\cos^2\theta + 2(\alpha - 1)}
	\right).
\end{equation}
In solving the inequalities (\ref{ineq:D2}) and (\ref{ineq:D2nr}), it is convenient to consider the cases $\alpha\le 1$ and $\alpha>1$. As a result, the solution involves two different regions:
\begin{equation}
	\label{sol:D1}
	\left\{
	\begin{array}{rcccl}
		-1 &\le& \cos\theta &\le& 1\\
		0  &\le& |\bm p| &\le& p_+(\cos\theta) \\
	\end{array}
	\right.,\ \text{for}\ \alpha > 1,
\end{equation}
and
\begin{equation}
	\label{sol:D2}
	\left\{
	\begin{array}{rcccl}
		c  &\le& {\cos\theta} &\le& 1 \\
		p_-(\cos\theta) &\le& |\bm p| &\le& p_+(\cos\theta)  \\
	\end{array}
	\right.,\ \text{for}\ \alpha_0 \le \alpha \le 1.
\end{equation}
The parameters $\alpha_0$ and $c$ are different for \Eqs{ineq:D2}{ineq:D2nr}.
For relativistic kinematics
$c=\sqrt{1-\alpha^2}/\beta$ and $p_\pm(\cos\theta)$ are given by Eq.(\ref{p:bounds:D}), while
for the nonrelativistic spectator, $c=\sqrt{2(1-\alpha)}/\beta$ and $p_\pm(\cos\theta)$ are given by Eq.(\ref{p:bounds:D:nr}).
The minimum value of $\alpha$ in \eq{sol:D2} is derived from the condition $c=1$.
We have $\alpha_0=\sqrt{1-\beta^2}$ and $\alpha_0=1-\tfrac12\beta^2$ for \Eqs{ineq:D2}{ineq:D2nr}, respectively.

Also, the condition $c=1$ determines the maximum allowed value of $x_d$ for given $Q^2$ which is consistent with \eq{eq:w2ineq1}.
For $Q^2\to\infty$, we have $x_d^\text{max}=1$ and $x_d^\text{max}=1-M/(2M_d)\approx 3/4$ for the relativistic and nonrelativistic spectator kinematics, respectively.
For finite values of $Q^2$ we have in case of \eq{ineq:D2},
\begin{equation}\label{eq:xdmax}
x_d^\text{max}=\left(1+((M+W_\text{th})^2-M_d^2)/Q^2\right)^{-1}
\approx \left(1+4m_\pi M/Q^2\right)^{-1}.
\end{equation}

We use \Eqs{sol:D1}{sol:D2} to cast the momentum integral in \eq{eq:IA} as follows:
\begin{equation}
	\label{int:D:sph}
	\int\ud^3\bm{p}\left|\Psi_d(\bm{p})\right|^2\theta\left(W^2{-}W_\text{th}^2\right) =
	\left\{
	\begin{aligned}
		&\frac12
		\int_{-1}^1\hspace{-0.5em}\ud\cos\theta
		\int_0^{p_+(\cos\theta)}\ud p\,
		p^2\left(\psi_0^2(p)+\psi_2^2(p)\right), & \text{for}\ \alpha>1,
\\
		&\frac12
		\int_{c}^1\hspace{-0.5em}\ud\cos\theta
		\int_{p_-(\cos\theta)}^{p_+(\cos\theta)}\ud p\,
		p^2\left(\psi_0^2(p)+\psi_2^2(p)\right), & \text{for}\ \alpha \le 1,
\\
	\end{aligned}
	\right.
\end{equation}
where $p=|\bm p|$, and $\psi_0$ and $\psi_2$ are the deuteron orbital momentum wave functions for $l=0$ and $l=2$, respectively:
\begin{equation}\label{eq:psid}
	\left|\Psi_d(\bm p)\right|^2=\left(\psi_0^2(p)+\psi_2^2(p)\right)/(4\pi).
\end{equation}
Following \eq{eq:wfnorm}, the functions $\psi_0$ and $\psi_2$ are normalized as follows:
\begin{equation}\label{eq:wfnorm2}
	\int_0^{\infty} \ud p\,p^2 \left(\psi_0^2+\psi_2^2\right)=1.
\end{equation}

\begin{figure}[htb]
\centering
\includegraphics[width=0.33\textwidth]{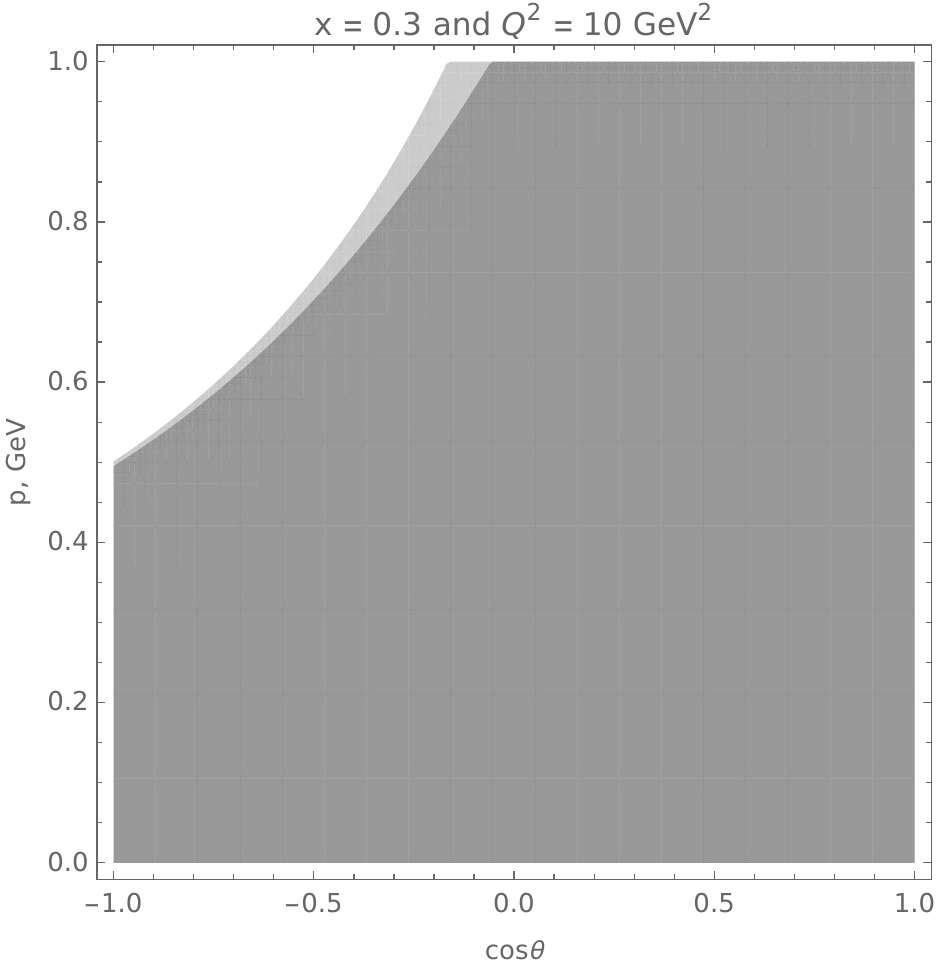}%
\includegraphics[width=0.33\textwidth]{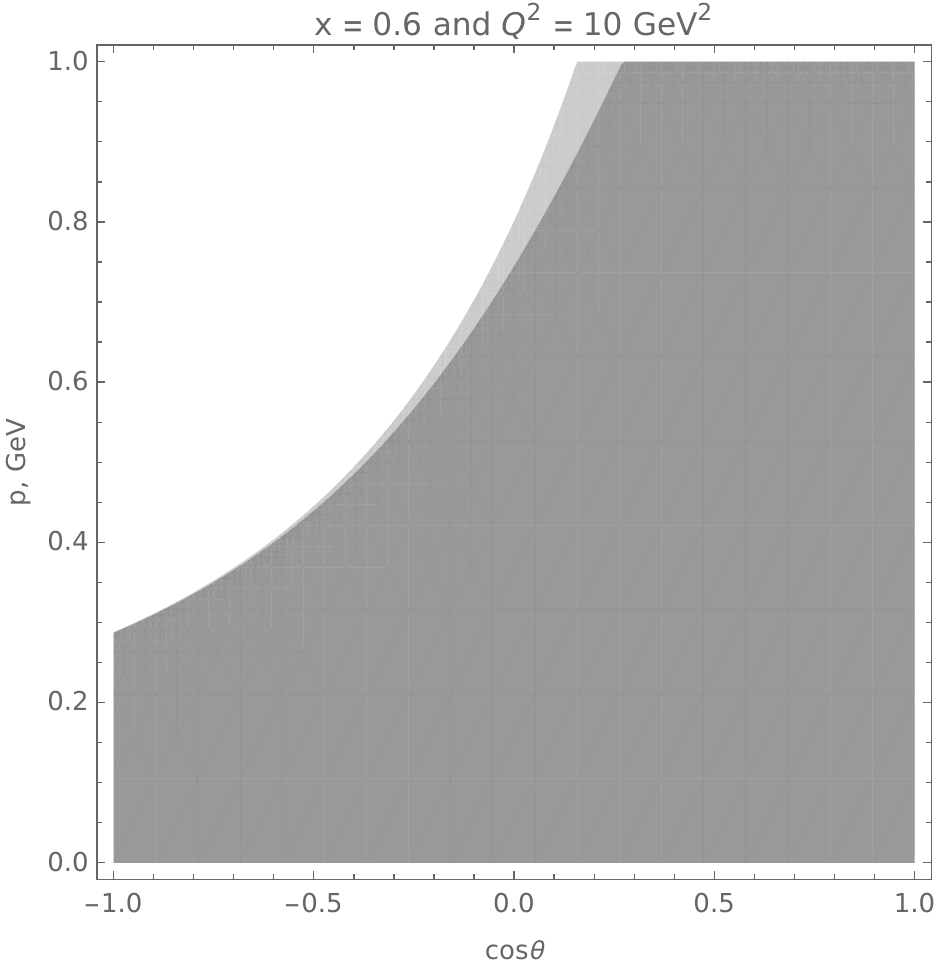}%
\includegraphics[width=0.33\textwidth]{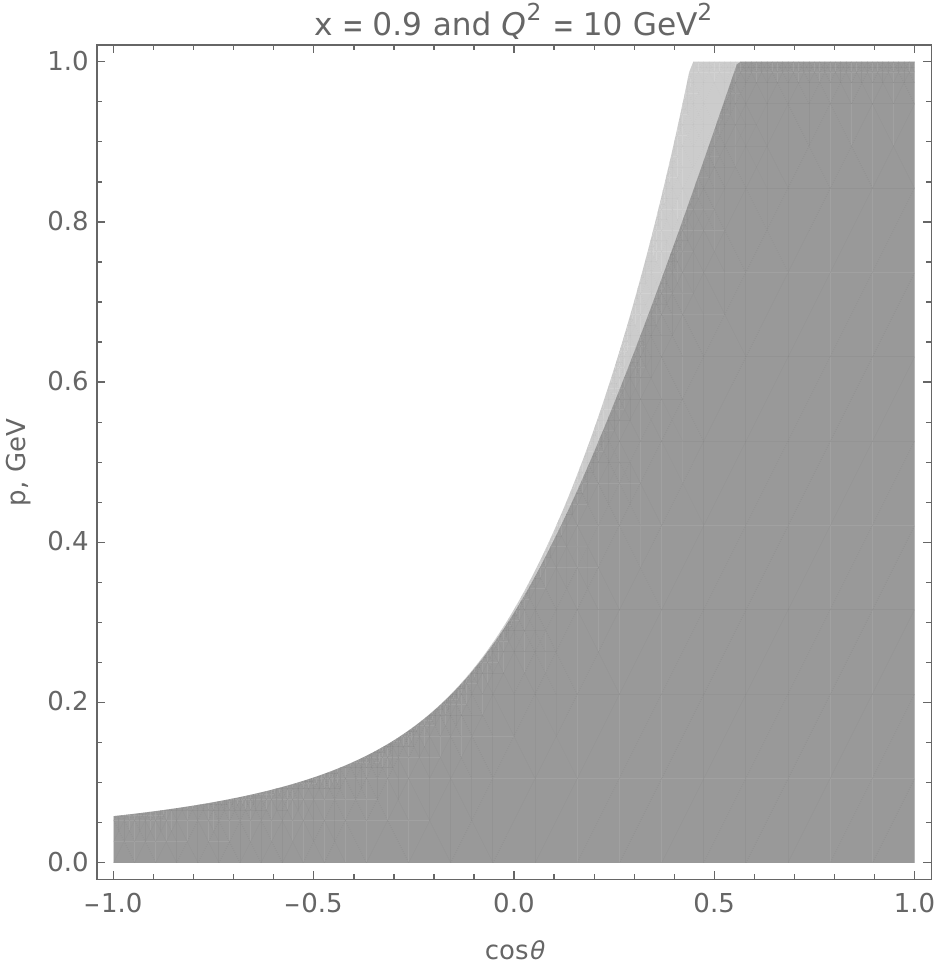}
\includegraphics[width=0.33\textwidth]{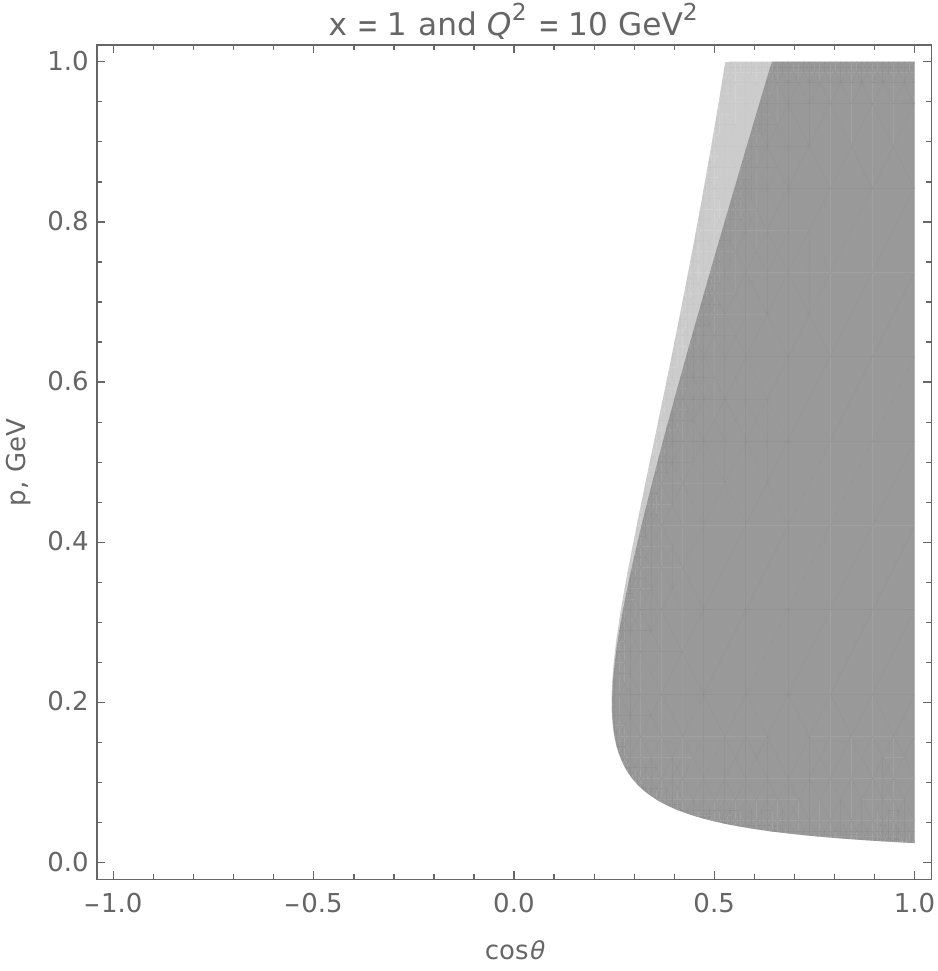}%
\includegraphics[width=0.33\textwidth]{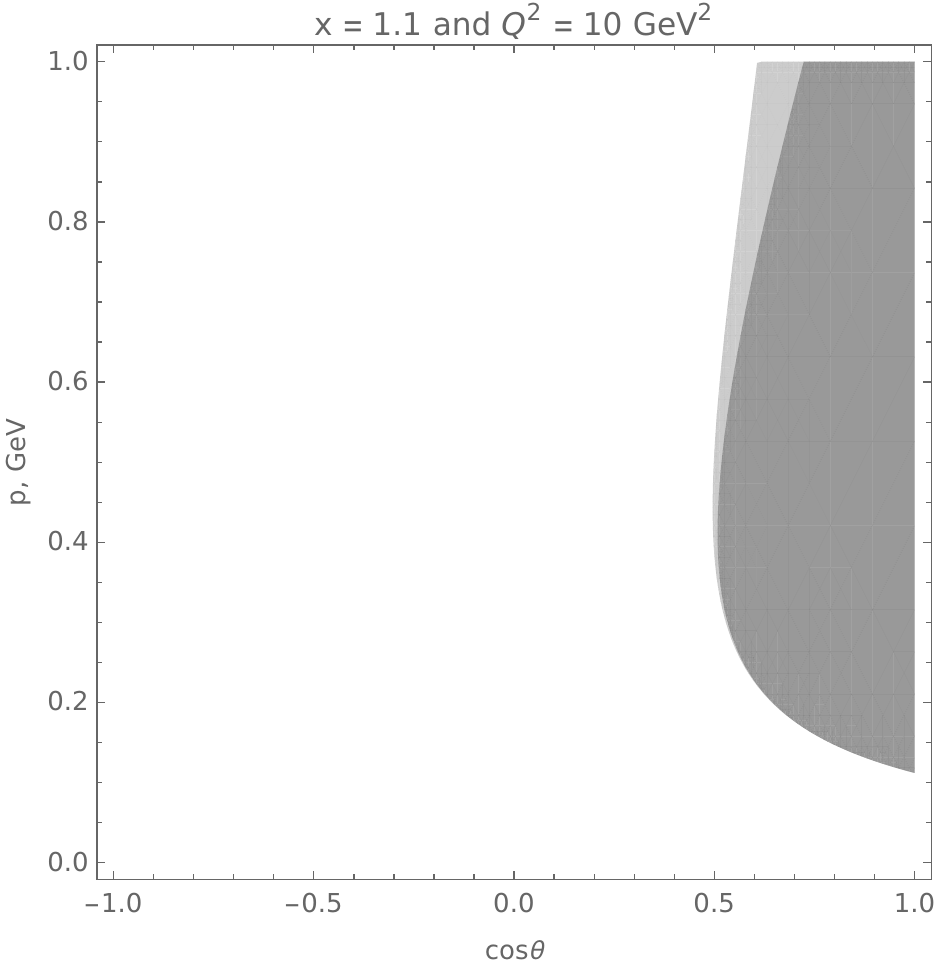}%
\includegraphics[width=0.33\textwidth]{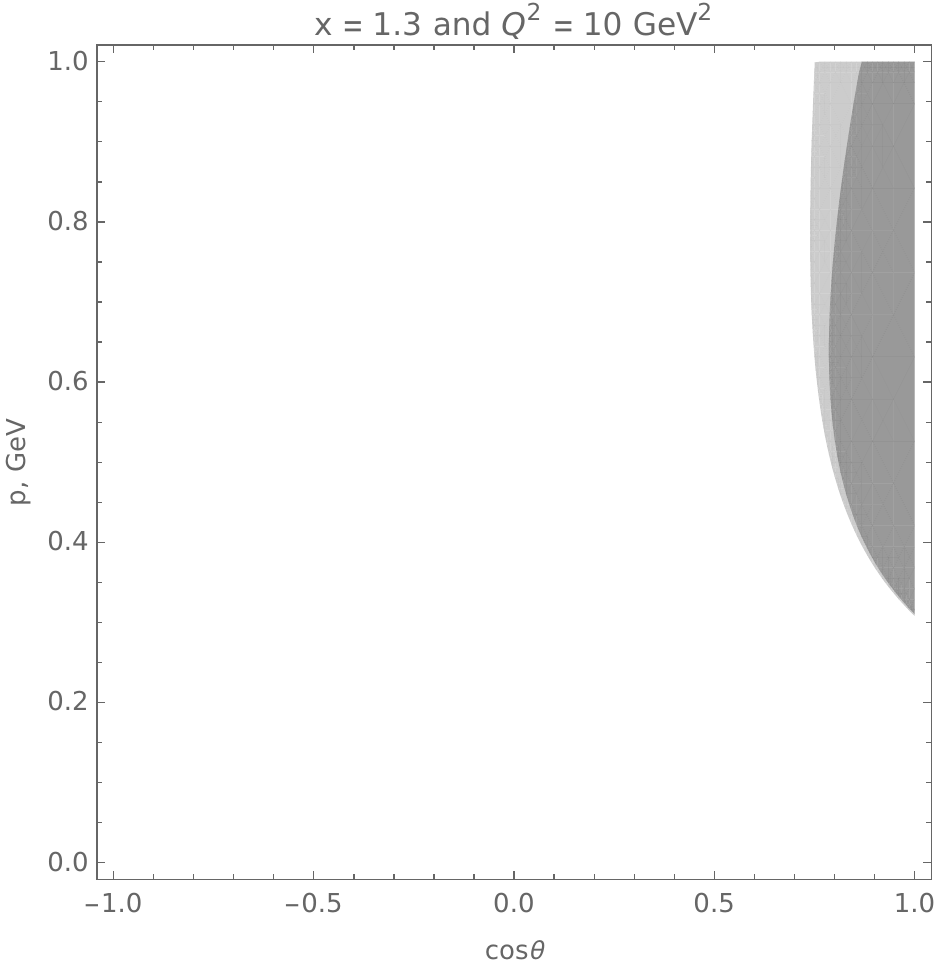}
\caption{%
The shaded area shows the region constrained by \Eqs{sol:D1}{sol:D2} together with the cut $p<1\gev$ for a few fixed values of $x$ and $Q^2$ shown in the panels. The light-gray region corresponds to a relativistic spectator, while the dark-gray is for the nonrelativistic one.
}
\label{fig:ps1}
\end{figure}

Note that in numerical applications, we apply a cut on the bound nucleon momentum $p_\text{cut}\sim 1\gev$ in the nuclear convolution.
For this reason, we replace the upper limit on the momentum in \eq{int:D:sph} with $\min(p_\text{cut},p_+(\cos\theta))$.
The integration region in \eq{int:D:sph} is illustrated in Fig.~\ref{fig:ps1} for
both the relativistic and the nonrelativistic spectator and
for a few fixed values of $x$ and $Q^2$.
The integration region systematically shrinks with rising $x$, and
the allowed kinematical region is somewhat larger for the relativistic case,
although the difference is only visible for high nucleon momenta $p>0.5\gev$.
As a somewhat extreme example of the deuteron kinematics,
in the last panel of Fig.~\ref{fig:ps1}, we show the integration region for $x=1.3$,
which is limited to high values of $\cos\theta$ and momentum $p>300\mev$.

\subsection{Convolution integral using $(y, p_\perp)$ basis}
\label{sec:conv:pty}

Consider \eq{eq:IA} and
note that $x'=x/y$, where the dimensionless variable $y=p\cdot q/(Mq_0)=(p_0+\gamma p_z)/M$
is usually referred to as the nucleon light-cone momentum.
The integral over the nucleon momentum in \eq{eq:IA} can be cast in terms of integration
over $y$ and $p_\perp$:
\begin{equation}\label{eq:conv:pty}
	F_i^d(x,Q^2) = \int\ud y\ud p_\perp^2 d_{ij}(y,p_\perp^2,\gamma) F_j^N(x/y,Q^2,\mu^2),
\end{equation}
where $i=T,2$ and we assume the sum over the repeated subscript $j=T,2$ and
\begin{align}\label{eq:dypt}
	d_{ij}(y,p_\perp^2,\gamma) &= \pi \int \ud p_z  \left|\Psi_d(\bm p)\right|^2 K_{ij}
	\delta\left(y-\frac{p_0 + \gamma p_z}{M}\right),
\end{align}
and $\mu^2=p_0^2-\bm p^2$, and $p_0$ is the energy of the active nucleon, and
the kinematical factors $K_{ij}$ in \eq{eq:conv:pty} are given by \eq{eq:Kij}.

Note that the kernel $d_{ij}$ in the convolution integral \eq{eq:conv:pty} depends on the kinematic variables $x$ and $Q^2$ through a dimensionless parameter $\gamma=(1+4x^2M^2/Q^2)^{1/2}$.
We now briefly consider the $\gamma=1$ case corresponding to the light-cone kinematics of $Q^2\to\infty$.
In this limit, the matrix $K_{ij}$ has only the diagonal components, and $K_{TT}=K_{22}=1+p_z/M$.
Then the nuclear convolution for $F_T$ and $F_2$ have the same form with the kernel $d_{TT}=d_{22}=D(y,p_\perp^2)$, which has the meaning of distribution over the corresponding variables:
\begin{equation}\label{eq:dypt1}
	D(y,p_\perp^2) = \pi \int \ud p_z  \left|\Psi_d(\bm p)\right|^2 \left(1+\frac{p_z}{M}\right)
	\delta\left(y-\frac{p_0 + p_z}{M}\right) .
\end{equation}
The distribution by \eq{eq:dypt1} is normalized to 1:
\begin{equation}\label{eq:dnorm}
\int \ud y\ud p_\perp^2 D(y,p_\perp^2)=\int \ud^3\bm p \left|\Psi_d(\bm p)\right|^2 \left(1+\frac{p_z}{M}\right) = 1.
\end{equation}
The term proportional to $p_z$ vanishes after angular integration.

Note that in the off-shell region, the nucleon SF in \eq{eq:conv:pty} depends on the virtual nucleon mass square $\mu^2$.
We use \eq{SF:OS} in order to separate the off-shell dependence of the bound nucleon structure function  and integrate over $\ptsq$. Then \eq{eq:conv:pty} can be cast in terms of a one-dimensional convolution integral as follows:
\begin{equation}\label{eq:conv:pty1}
	F_2^d(x,Q^2) = \int\limits_{x}^{y_\text{max}}\ud y\left[
	S_0(y) F_2^N(x/y,Q^2) + S_1(y)\delta f(x/y) F_2^N(x/y,Q^2)
	\right],
\end{equation}
where the light-cone smearing functions $S_0$ and $S_1$ are as follows:
\begin{align}
\label{eq:s0}
S_0(y) &= \int\ud\ptsq D(y,\ptsq),
\\
\label{eq:s1}
S_1(y) &= \int\ud\ptsq D(y,\ptsq)v,
\end{align}
where $v=(\mu^2-M^2)/M^2$ is the nucleon virtuality. The function $S_0(y)$ makes sense of the nucleon light-cone distribution in the deuteron and normalized to unity according to \eq{eq:dnorm}.
Note that \eq{eq:conv:pty1} was derived for $\gamma=1$, i.e., light-cone kinematics. In this limit, the constraint by \eq{eq:w2ineq1} reduces to $y>x$ and $y_\text{max}=M_d/M$.

Below we discuss in more detail the nuclear convolution by \Eqs{eq:conv:pty}{eq:dypt} for both the relativistic and nonrelativistic kinematics of the nucleon spectator while keeping finite $Q^2$ effects.

\subsubsection{Nonrelativistic spectator}
\label{sec:nr}

\begin{figure}[htb]
\includegraphics[width=0.33\textwidth]{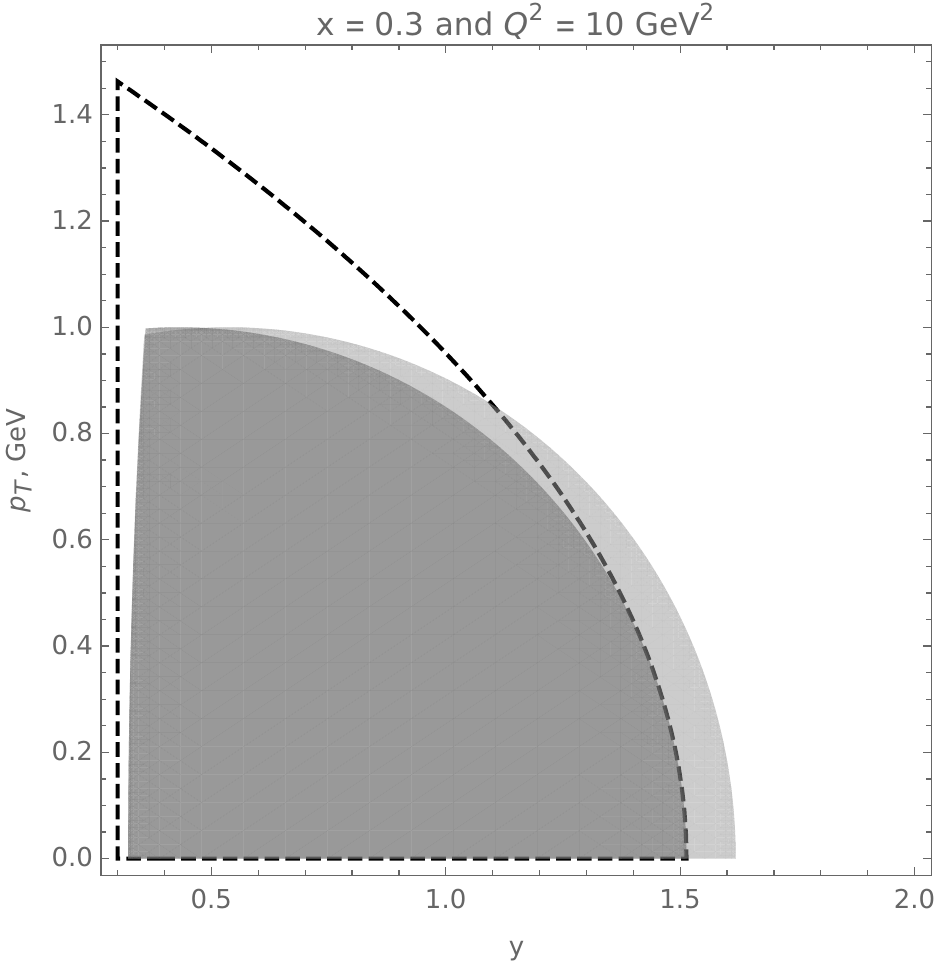}%
\includegraphics[width=0.33\textwidth]{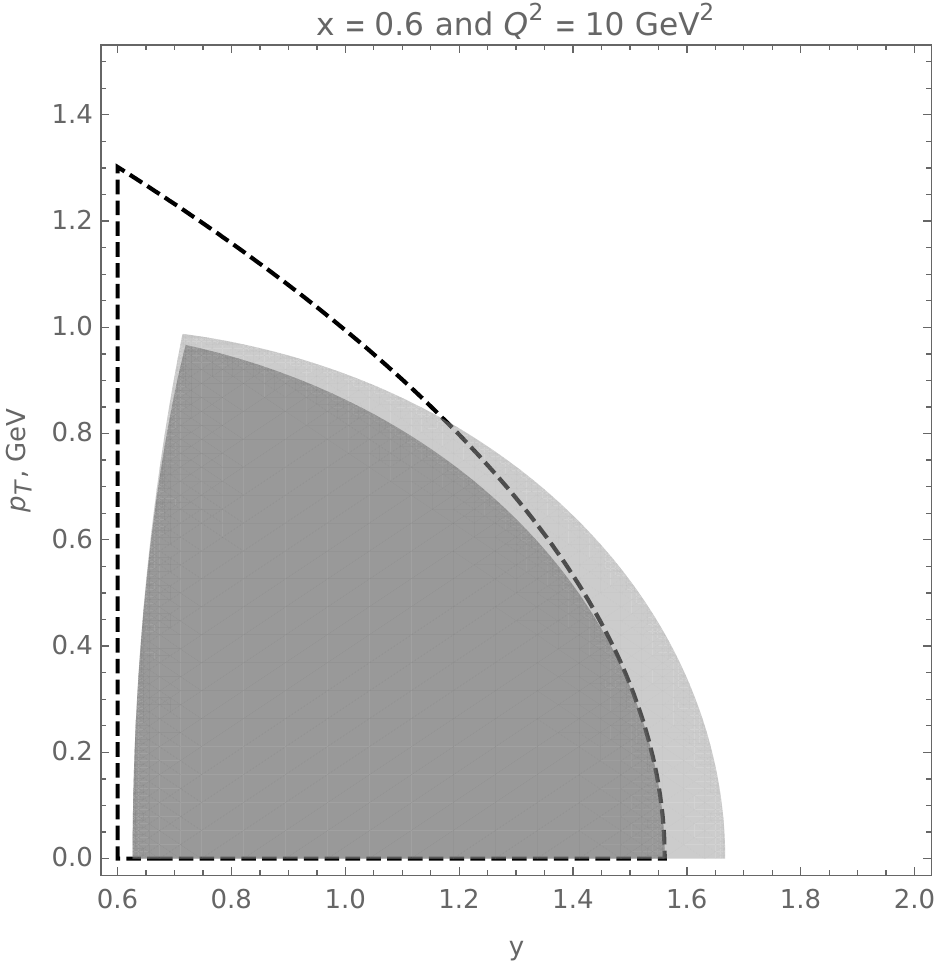}%
\includegraphics[width=0.33\textwidth]{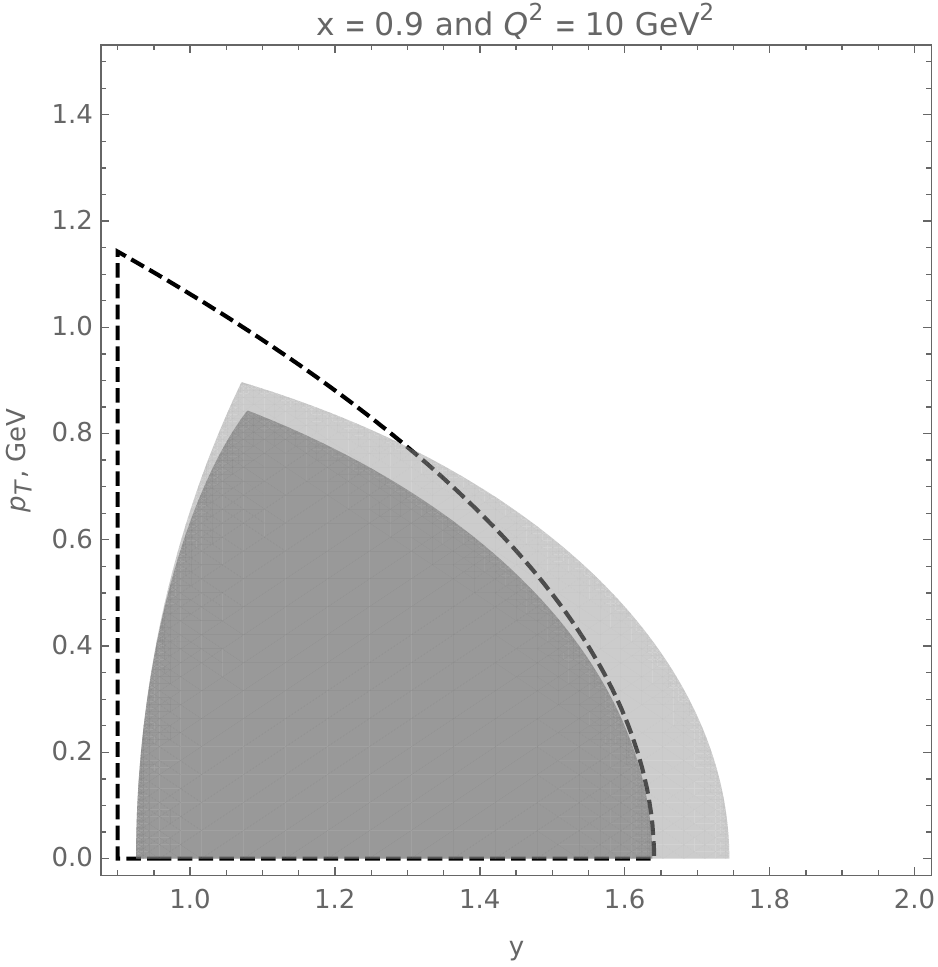}
\includegraphics[width=0.33\textwidth]{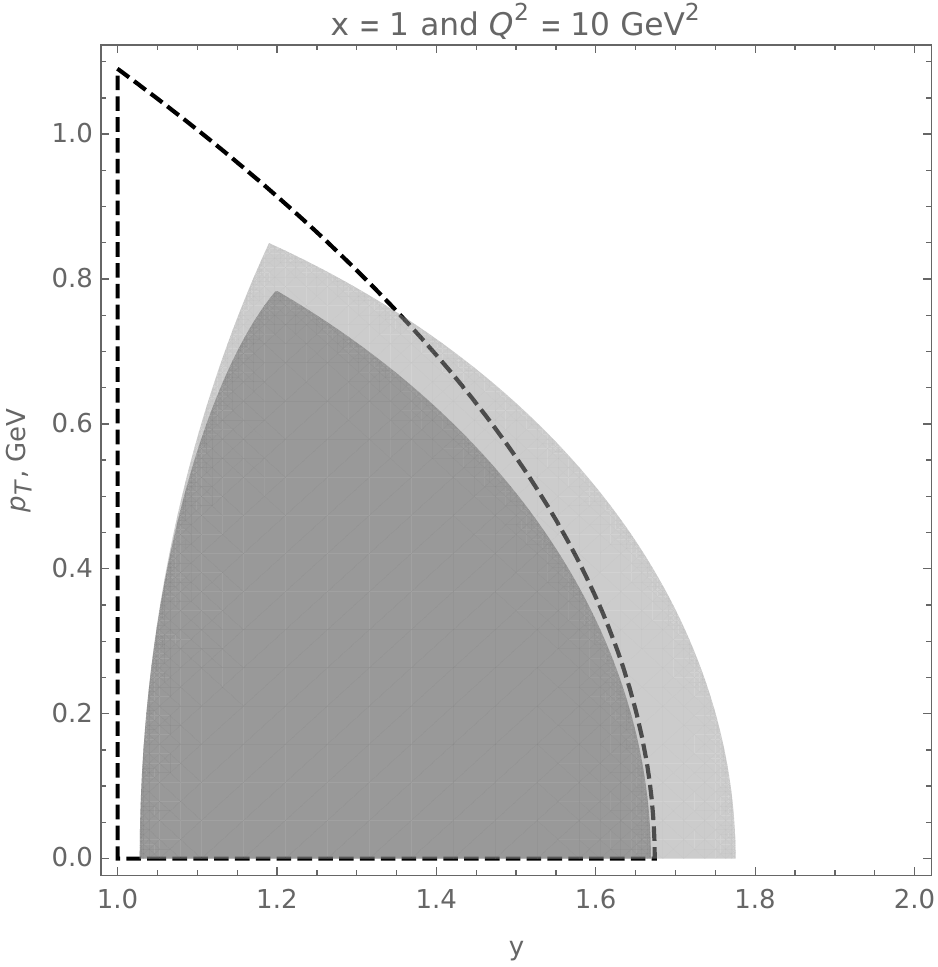}%
\includegraphics[width=0.33\textwidth]{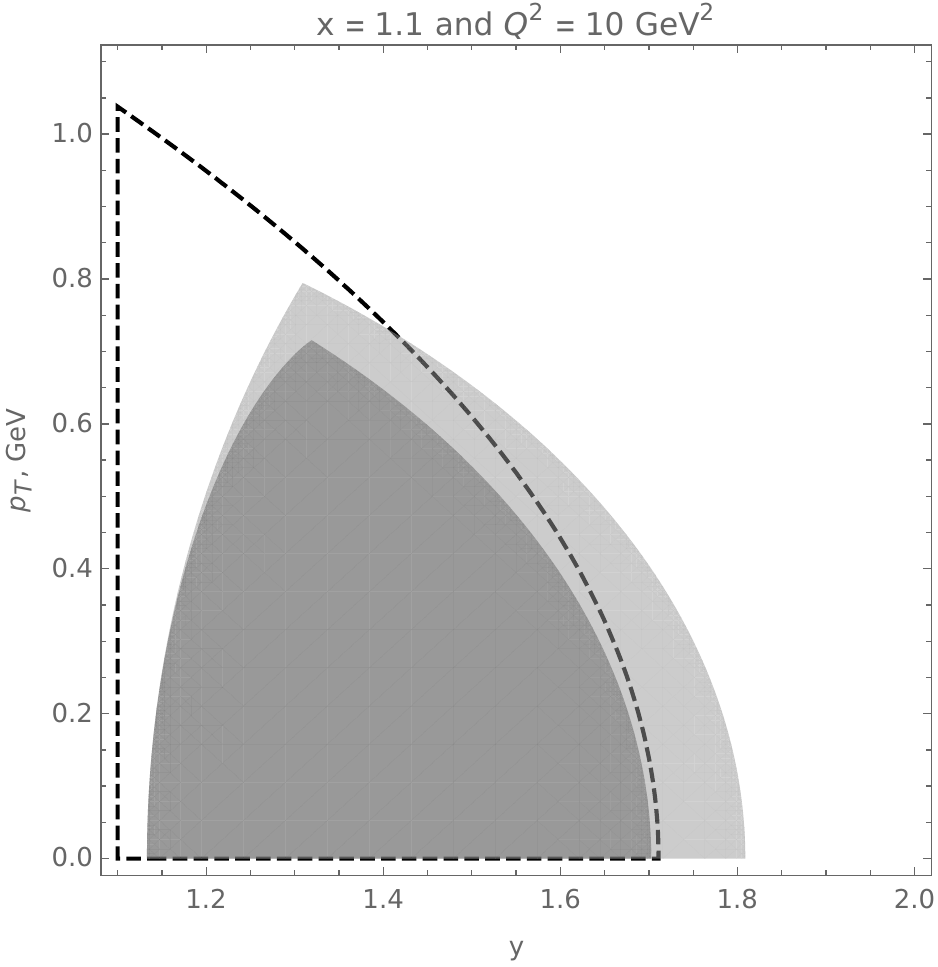}%
\includegraphics[width=0.33\textwidth]{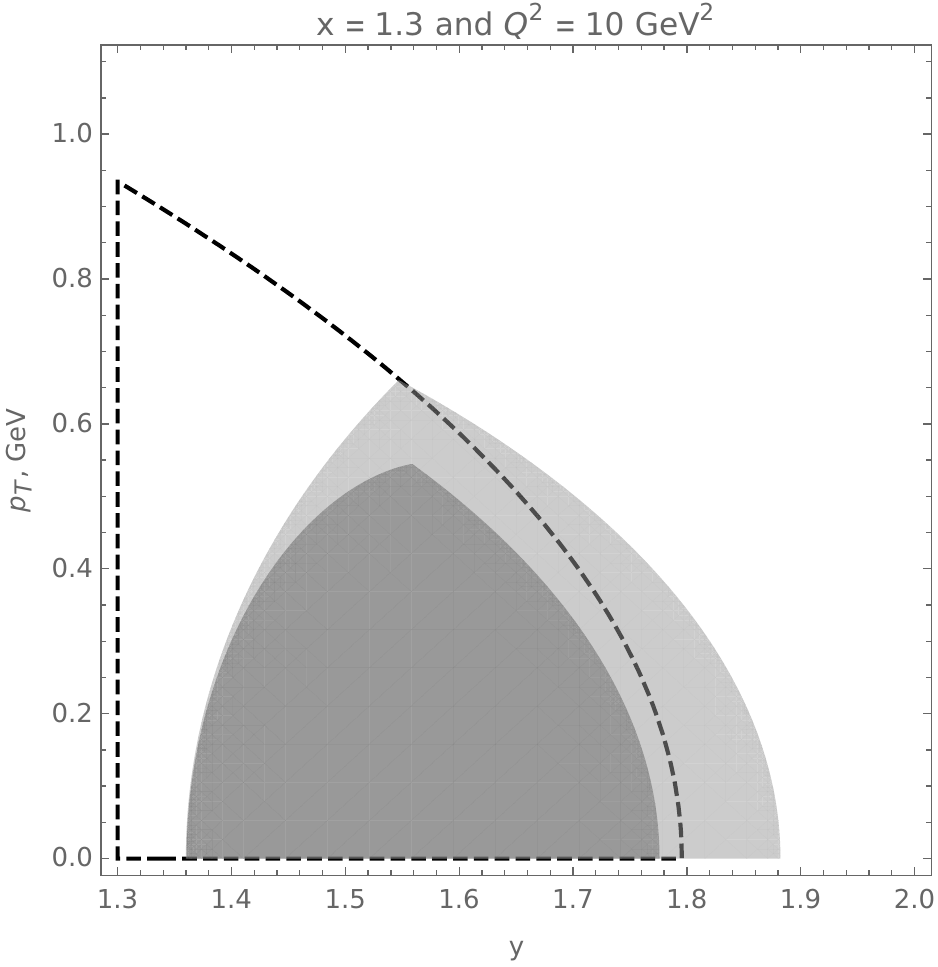}
\caption{%
The integration region in \eq{conv:pty:nr} for different values of $x$ and $Q^2$.
The region $x<y<y_\text{max}$ and $p_\perp^2<t^2(y)$ is shown by dashed lines.
The shaded area is a region restricted to $W^2>(M+m_\pi)^2$ and $|\bm p|<1\gev$
computed for both the relativistic and the nonrelativistic spectator,
and for fixed values of $x$ and $Q^2$ indicated in the panels.
The light-gray region corresponds to the relativistic spectator,
while the dark-gray region is for the nonrelativistic one.}
\label{fig:ps:ypt}
\end{figure}

We first consider  \eq{eq:dypt} assuming the \emph{nonrelativistic}
nucleon with energy $p_0=M+\ceps_d-\bm p^2/(2M)$,
where $\ceps_d=M_d-2M$ is the deuteron binding energy.
Taking the integral in \eq{eq:dypt} we have
\begin{equation}\label{eq:intpz:nr}
	\int \ud p_z \delta\left(y-\frac{p_0 + \gamma p_z}{M}\right) = \frac{M^2}{\sqrt{t^2-p_\perp^2}},
\end{equation}
where
\begin{align}\label{eq:t:nr}
	t^2 &= 2M^2 \left(y_\text{max} - y\right),
\\
	y_\text{max} &= 1+\frac{\gamma^2}{2} + \frac{\ceps_d}{M} .
\label{eq:ymax:nr}
\end{align}
Note also that by integrating the $\delta$ function in \eq{eq:intpz:nr},
we have $p_z$ as a function of $y$ and $p_\perp^2$:
\begin{align}\label{eq:pz:nr}
	p_z = \gamma M - \sqrt{t^2-p_\perp^2}.
\end{align}
Note that $t$ makes sense of the maximum $p_\perp$ for the given $y$.
The condition $t^2=0$ determines the maximum value $y=y_\text{max}$; see  \eq{eq:ymax:nr}.
Note that for $\gamma=1$ and neglecting a small correction due to the deuteron binding energy in \eq{eq:ymax:nr}, we have
$y_\text{max}=3/2$. This is different from  the kinematical maximum $y_\text{max}=M_d/M$
in the relativistic case which is discussed in Sec.~\ref{sec:rel} and illustrated in Fig.~\ref{fig:ps:ypt}.

Using \eq{eq:intpz:nr} to (\ref{eq:pz:nr}), we cast \eq{eq:conv:pty} as follows:
\begin{align}\label{conv:pty:nr}
	F_i^d(x,Q^2) = \frac{M^2}{4} \int\limits_{y_\text{min}}^{y_\text{max}}
	\ud y \int\limits_0^{t^2}\ud p_\perp^2
	\frac{\left(\psi_0^2(p)+\psi_2^2(p)\right)}{\sqrt{t^2-p_\perp^2}}
	K_{ij} F_j^N(x/y,Q^2,\mu^2) ,
\end{align}
where $p=\sqrt{p_z^2+\ptsq}$ and $p_z$ is given by \eq{eq:pz:nr}.
Note the $\ptsq$ integration in \eq{conv:pty:nr} has a singularity at  $\ptsq=t^2$.
Although this is an integrable singularity, it may cause an instability in numerical applications.
For this reason, it is convenient to change the integration variable in \eq{conv:pty:nr}
from $p_\perp^2$ to $u=\sqrt{t^2-p_\perp^2}$.
Then we have
\begin{align}\label{conv:pty:nr2}
	F_i^d(x,Q^2) = \frac{M^2}{2} \int\limits_{y_\text{min}}^{y_\text{max}}
	\ud y \int\limits_0^{t}\ud u
	\left(\psi_0^2(p)+\psi_2^2(p)\right)
	K_{ij} F_j^N(x/y,Q^2,\mu^2),
\end{align}
where $p_\perp^2=t^2-u^2$, $p_z=\gamma M - u$, $p=\sqrt{p_\perp^2+p_z^2}$, and $\mu^2=M^2+2M\ceps_d-2p^2$.
The lower limit of integration over the light-cone variable is $y_\text{min}=x$.
Note, however, that this integration region in the nuclear convolution is modified for finite $Q^2$.
The corresponding region can be inferred from \eq{eq:w2ineq1}.
Unlike the case of spherical coordinates discussed in Sec.~\ref{sec:conv2:spherical},
the analytic solution to the inequality (\ref{eq:w2ineq1}) in terms of $(y,p_\perp)$ is somewhat cumbersome and not shown here.
The resulting integration region in the nuclear convolution \eq{conv:pty:nr} is illustrated in Fig.~\ref{fig:ps:ypt}.

\begin{figure}[htb]
\centering
\includegraphics[width=0.5\textwidth]{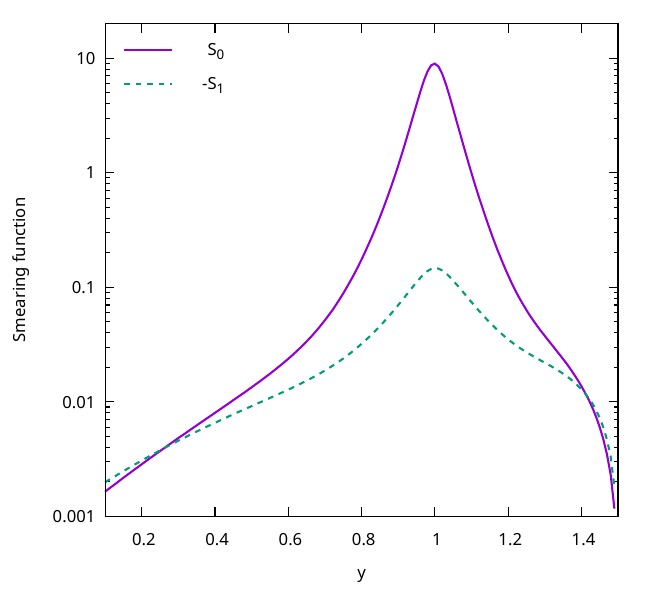}%
\includegraphics[width=0.5\textwidth]{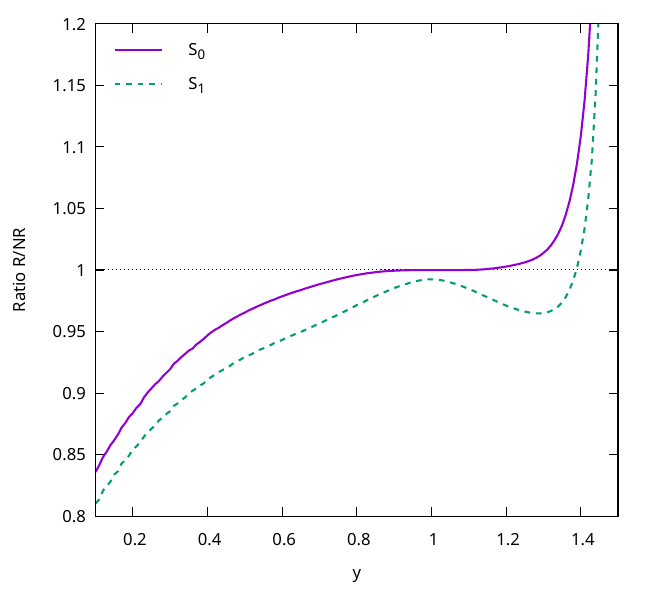}
\hfill
\caption{\label{fig:s01}
Left panel shows the smearing functions $S_0(y)$ (solid line) and $-S_1(y)$ (dashed line)
computed using \Eqs{eq:s0:nr}{eq:s1:nr} for the AV18 deuteron wave function.
The right panel illustrates the relativistic effects in the smearing functions (see also text).
}
\end{figure}

In conclusion of this section we present the explicit expressions for the light-cone smearing functions by \Eqs{eq:s0}{eq:s1}:
\begin{align}
\label{eq:s0:nr}
S_0(y) &= \frac{M^2}{2}\int\limits_0^{t}\ud u\left(\psi_0^2(p)+\psi_2^2(p)\right)\left(2-\frac{u}{M}\right),
\\
\label{eq:s1:nr}
S_1(y) &= \frac{M^2}{2}\int\limits_0^{t}\ud u\left(\psi_0^2(p)+\psi_2^2(p)\right)\left(2-\frac{u}{M}\right) v ,
\end{align}
where  $v=\mu^2/M^2-1$ and the other notations  are similar to those in \eq{conv:pty:nr2}.
The function $S_{0}(y)$ and $S_{1}(y)$ computed for the AV18 deuteron wave function are plotted in Fig.\ref{fig:s01} (left panel).

\subsubsection{Relativistic spectator}
\label{sec:rel}

For the \emph{relativistic} kinematics, we have $p_0=M_d-\sqrt{M^2+\bm p^2}$.
Integrating the $\delta$ function in \eq{eq:dypt}, we have
\begin{equation}\label{eq:intpz:r}
	\int \ud p_z \delta\left(y-\frac{p_0 + \gamma p_z}{M}\right) =
	\frac{\gamma M E}{aM + (\gamma^2-1)E},
\end{equation}
where $a=M_d/M-y$, $E=\sqrt{p_z^2 + p_\perp^2 + M^2}$, and we should replace $p_z$ with the solution
of the following equation:
\begin{equation}\label{eq:pz:r}
	\gamma p_z = \sqrt{p_z^2 + p_\perp^2 + M^2} - aM .
\end{equation}
For $\gamma>1$, this equation has a solution for any value of the parameter $a$:
\begin{equation}\label{eq:pz:r:sol}
	p_z = \frac{-\gamma a M +\sqrt{a^2M^2 +(\gamma^2-1)(M^2+p_\perp^2)}}{\gamma^2-1}.
\end{equation}
Note, that for $\gamma\to1$ (or $Q^2\to\infty$), only the region $a>0$ is allowed, and we have
\begin{equation}\label{eq:pz:r1}
	p_z = \frac{p_\perp^2 + (1-a^2)M^2}{2aM}.
\end{equation}
In this case, the condition $a=0$ determines the upper limit on $y$, $y_\text{max}=M_d/M$.
For finite values of $Q^2$, the integration region extends to $y>M_d/M$.

For the deuteron structure functions, we have
\begin{align}\label{conv:pty:r}
	F_i^d(x,Q^2) = \frac{\gamma}{4} \int\limits_{y_\text{min}}^{y_\text{max}} \ud y
	\int \ud p_\perp^2
	\frac{M E}{a M + (\gamma^2-1)E} \left(\psi_0^2(p)+\psi_2^2(p)\right)
	K_{ij} F_j^N(\frac{x}{y},Q^2,\mu^2).
\end{align}
The integration region in \eq{conv:pty:r} is limited by \eq{eq:w2ineq1}.
The resulting region is shown in Fig.\ref{fig:ps:ypt},
in which we also illustrate the impact of the momentum cut on the integration region.

The light-cone distributions by \Eqs{eq:s0}{eq:s1} can be written as
\begin{align}
\label{eq:s0:r}
S_0(y) &= \frac{1}{4a} \int\ud p_\perp^2 \left(\psi_0^2(p)+\psi_2^2(p)\right) E\left(1+\frac{p_z}{M}\right),
\\
\label{eq:s1:r}
S_1(y) &= \frac{1}{4a} \int\ud p_\perp^2 \left(\psi_0^2(p)+\psi_2^2(p)\right) E\left(1+\frac{p_z}{M}\right) \frac{M_d^2-2M_d E}{M^2} .
\end{align}
Note that these functions have a pole at $a=0$ that corresponds to $y=M_d/M$.
However, this value of $y$ requires an infinite nucleon momentum as $p_z\to\infty$ at $y= M_d/M$.
Such configurations  should be suppressed by the deuteron wave function.
In practice, the region of large $y\sim M_d/M$, and therefore the singularity, can be avoided
by applying a reasonable cut on the nucleon momentum in the convolution integral.

The effect of relativistic kinematics is illustrated in Fig.~\ref{fig:s01}
(right panel),
in which we show the ratio of the function $S_0$ computed with \Eqs{eq:s0:r}{eq:s0:nr}
and a similar ratio for $S_1$.
For the most important region $|y-1|<0.2$, which drives the nuclear convolution,
this relativistic effect is negligible for $S_0$.
For this reason the relativistic effect has only a small impact on the deuteron structure function for $x<1$.
The relativistic correction is somewhat larger for $S_1$
but does not exceed $3\%$ in this region.
The region of large $|y-1|$ is driven by a high-momentum component
of the deuteron wave function.
For this reason, the effect of relativistic kinematics
on the smearing functions is more important in this region,
as illustrated by Fig.~\ref{fig:s01}.
As the region $y>1$ drives the deuteron structure functions for $x\gtrsim 1$,
one cannot ignore the effect of relativistic kinematics in this region of $x$.

\subsection{Benchmarks of the convolution integral}
\label{sec:bench}

In order to facilitate the comparison with the present approach, in Table~\ref{tab:bench} we list our results for $F_2^d$ computed for the test functions $F_2^N=(1-x)^3$ and $\delta f=x$.

\begin{table}[htb]
    \centering
\caption{The values of the deuteron structure function $F_2^d$ computed by \eq{eq:IA} using the AV18 deuteron wave function~\cite{Veerasamy:2011ak} and the test functions $F_2^N=(1-x)^3$ and $\delta f=x$.}
    \label{tab:bench}
\begin{tabular*}{\textwidth}{|l@{\extracolsep{\fill}}|c|c|c|c|c|c|c|} \hline
\diagbox{$Q^2$}{$x$} & 0     & 0.2       & 0.4       & 0.6       & 0.8       & 0.9       & 1.0\\ \hline
   1.0               & 1.000 & 5.032E-01 & 2.086E-01 & 6.188E-02 & 9.986E-03 & 2.977E-03 & 8.851E-04\\ \hline
  10.0               & 1.000 & 5.034E-01 & 2.095E-01 & 6.198E-02 & 8.711E-03 & 1.785E-03 & 2.359E-04\\ \hline
 100.0               & 1.000 & 5.035E-01 & 2.096E-01 & 6.198E-02 & 8.550E-03 & 1.631E-03 & 1.649E-04\\ \hline
\end{tabular*}
\end{table}

\bibliography{literatura}

\end{document}